\documentclass[preprint]{aastex}
\usepackage{natbib}
\usepackage{longtable}

\shorttitle{Stellar populations of HCG galaxies}
\shortauthors{Mendes de Oliveira et al.}

\received{2004 April 30}
\begin{document}

\title{Ages, metallicities and \\
$\alpha$-element enhancement for  galaxies in  Hickson compact groups }

\author{C. Mendes de Oliveira \altaffilmark{1,2,3}, P. Coelho
\altaffilmark{1,4}, J.J. Gonz\'alez \altaffilmark{5} and B. Barbuy
\altaffilmark{1}}

\altaffiltext{1}{Universidade de S\~ao Paulo, IAG,
Departamento de Astronomia, Rua do Mat\~ao 1226, Cidade
Universit\'aria, 05508-900, S\~ao Paulo, SP, Brazil, oliveira@astro.iag.usp.br, pcoelho@usp.br, barbuy@astro.iag.usp.br}

\altaffiltext{2}{Max-Planck-Institut f\"ur Extraterrestrische Physik,
Giessenbachstrasse, D-85748 Garching b. M\"unchen, Germany, oliveira@mpe.mpg.de}

\altaffiltext{3}{Universit\"ats-Sternwarte der Ludwig-Maximilians-Universit\"at,
Scheinerstrasse 1, D-81679 M\"unchen, Germany, oliveira@usm.uni-muenchen.de}

\altaffiltext{4}{Max-Planck-Institut f\"ur Astrophysik,
Karl-Schwarzschild-Str. 1, D-85741 Garching b. M\"unchen, Germany,
pcoelho@mpa-garching.mpg.de}

\altaffiltext{5}{Instituto de Astronomia, UNAM, Apdo Postal 70-726, 04510 Mexico D.F., Mexico}


% Mark off your abstract in the ``abstract'' environment. In the manuscript

\begin{abstract}

  Central velocity dispersions and eight line-strength Lick indices
have been determined from 1.3${\rm \AA}$ resolution long-slit spectra of
16 elliptical galaxies in Hickson compact groups. These data were used
to determine galaxy properties (ages, metallicities and $\alpha$-element
enhancements) and allowed a comparison with the parameters determined for
a sample of galaxies in lower density environments, studied by Gonz\'alez
(1993). The stellar population parameters were derived by comparison
to single stellar population models of Thomas et al. (2003) and
to a new set of SSP models for the indices Mg$_2$, Fe5270 and Fe5335 based
on synthetic spetra. These models, based on an update version of the fitting
functions presented in Barbuy et al. (2003), are fully described here.
Our main results are: (1) the two samples have similar mean
values for the metallicities and [$\alpha$/Fe] ratios, (2) the majority
of the galaxies in compact groups seem to be old (median age of 14 Gyr
for eight galaxies for which ages could be derived), in agreement with
recent work by Proctor et al. (2004).  These findings support two possible
scenarios: compact groups are either young systems whose members have
recently assembled and  had not enough time to experience any merging
yet or, instead, they are old systems that have avoided merging since
their time of formation.

\end{abstract}




\section{Introduction}

  The conventional scenario for the evolution of compact groups,
supported by n--body simulations, is that their members will interact
and merge into one single elliptical galaxy, if they are genuine
high-density systems.  Compact groups are, therefore, ideal environments in
which to study the formation of elliptical galaxies through mergers and
the effects of collisions on galactic evolution although this requires
that the lifetime of the group be longer than the merging timescales
between member galaxies.  We currently lack important information on
how long the member galaxies of a compact group have been together as
a compact entity although we have qualitative information that some
dynamical interaction among the galaxies has happened in the majority
of the groups.

While there is plenty of evidence that the rate of {\it
interactions} (which leave the two galaxies distinct) is high among
compact group galaxies, studies of optical colors and galaxy
morphologies suggest that the rate of ongoing mergers is low.  However,
it may be more difficult to recognize merger remnants in compact groups
than in other environments, since the forces exerted by the member
group galaxies tend to disrupt long tails and extensions created in
galactic collisions (Barnes \& Hernquist 1992). In addition, mergers in
compact groups may be different in nature if the colliding galaxies
have a smaller reservoir of gas. This may actually be the case since it
has been shown that at least cold gas seems to be depleted from small
galaxy groups, possibly due to interactions.

Several programs have been initiated to develop more sensitive tests of
merging history in elliptical galaxies in compact groups.  In this
context, a comparison of the 
stellar populations of elliptical galaxies
in compact groups with those of galaxies in other environments are
of great interest and may provide a quantitative measure of
the effects of mergers.

  A number of papers have studied the stellar populations
of early-type galaxies in clusters and in lower density environments
(e.g.  Rose et al. 2004; Jorgensen 1997; Bernardi et al. 1998;
Proctor et al. 2004). In compact groups, however, there are very few
previous studies.  Proctor et al. (2004) derived ages,
metallicities and $\alpha$-enhancements of stellar populations for a
sample of 17 elliptical galaxies and 9 spiral bulges in compact groups
and found that the HCG galaxies are generally older and more metal rich
than their field counterparts.  Although Jorgensen (1997) argues that it
is not possible to derive unique galaxy parameters from the observables,
some more recent studies, e.g.  Thomas et al. (2003, hereafter TMB03),
have developed stellar population models that include element abundance
ratio effects and which can provide a derivation of age, total metallicity
and element ratios from Lick absorption line indices. A complication is
however the degeneracy between age and horizontal branch morphology,
which comes from the fact that the presence of warm horizontal branch
stars may strengthen the Balmer absorption lines, possibly mimicing
a younger stellar population age (e.g. Maraston \& Thomas 2000). This
problem still remains to be solved.

The main contribution of this paper is the determination of internal
velocity dispersions and eight Lick/IDS indices for 16 galaxies in HCGs
(three of which are in common with the sample of Proctor et al. 2004).
Our new measurements were compared with models from TMB03, and with a new
set of SSP models, presented here for the first time, 
to derive ages, metallicities and
$\alpha$-element enhancements for the galaxies.

This paper is organized as follows. Section 2 describes our observations
and reductions. In Section 3 results from the comparisons between the
HCG and control sample regarding their ages, metallicities, and $\alpha$
enhancements are shown.  In addition, fitting functions
for the indices Mg$_2$, Fe5270 and Fe5335 based on synthetic spectra
(Barbuy et al. 2003), and a new set of SSP
model indices are presented.  In Section 4 we
discuss the possible selection effects affecting our sample, the main
contributions given by other work in the literature and the insights
provided by our new data regarding the merging histories of the compact
group galaxies.

\section{Observations and Reductions}

We observed 16 of the 28 Hickson compact
group (Hickson 1982) elliptical galaxies
south of declination --10$^\circ$, between R.A's 19$^h$ and 03$^{\circ}$, and
brighter than B$_T$ = 16 mag. 

    Long-slit spectra for the galaxies were taken at the Anglo
Australian Observatory with a 25cm camera, a 1200 V grating and a TEK CCD
on the AAT 3.5m telescope.  These spectra cover the range $\lambda\lambda$
$\sim$ 4810 -- 5630 ${\rm \AA}$ with 1.34 ${\rm \AA}$ resolution FWHM.
The wavelength and spatial scale were $\sim$ 0.79 ${\rm \AA}$/pixel and
1.03 arcsec/pixel respectively. Each galaxy was observed twice along
its major axis. The total exposure times ranged from 800s to 2000s. A
slit width of 1.5 arcsec was used.

 Radial velocities, internal velocity dispersions and the
Lick/IDS indices H${\beta}$, Mg$_1$, Mg$_2$, Mg$_b$, Fe5015, Fe5270,
Fe5335, Fe5406, inside an aperture of radius r$_{eff}$/2 
(where r$_{eff}$ is the effective
radius of the galaxy, obtained from the fit to a de~Vaucoleurs profile,
de~Vaucoleurs 1948) were obtained using the technique developed by
Gonz\'alez (1993).  In this method the spectra of each galaxy and 
template are compared in Fourier space. Optimal templates were 
constructed from an optimal linear combination of stellar
spectra
of late-type dwarf and giant stars obtained in the same nights when the
galaxy observations were made (see Gonz\'alez 1993, 
at http://www.astroscu.unam.mx/$\sim$jesus/, 
for the reduction
procedure details). Line-strength measurements were made in the Lick/IDS
system as introduced by Burstein et al. (1984) and refined by Worthey
et al. (1994). The transformation to the Lick/IDS system was done by
using stars from Faber et al. (1985)'s list observed during our run.

  We used a control sample taken from the
list of galaxies in loose groups and in the field studied by Gonz\'alez
(1993, hereafter the ``field'' sample), observed at Lick Observatory
with the 120-inch Shane Telescope, at a resolution of FWHM = 2.7 --
3.3 ${\rm \AA}$.  Velocity dispersions and line-strength indices for the
galaxies in the control sample were derived in the same manner as for
the galaxies in the compact group sample.  For comparison, two of the
elliptical galaxies in the control sample were observed in the run when
the spectra for the compact-group galaxies were obtained.  Their derived
velocity dispersion and line-strength indices are in excellent agreement
(within the quoted errors of 5\%) with those obtained independently by
Gonz\'alez (1993).

    All spectra were extracted inside a
radius of r$_{eff}$/2 in order to sample similar parts of the galaxies,
and avoiding the need of using uncertain aperture corrections.
This is important in our case since the control sample is composed of
galaxies which are quite nearby, with a median radial velocity of 2300
km s$^{-1}$, while the sample of compact group galaxies has a median
velocity of 7200 km s$^{-1}$.

\section{Results}

   The kinematic data for the Hickson compact group
galaxies are presented in Table 1. Radial velocities and velocity
dispersions were derived within an aperture of radius r$_{eff}/2$. The
effective radii and values for the surface brightness measured within the 
effective radii, listed in columns 2 and 3
were taken from Zepf and Whitmore (1993) for the galaxies in common and
were derived from B images taken at CFHT (Mendes de Oliveira 1992), for
the remaining galaxies.  In the fourth column, the S/N of the spectra
are listed. These were obtained within the spectral window 4900 to 5500
{\rm \AA}, derived similarly to those obtained by Gonz\'alez (1993) for the
control sample.  The Lick indices and their estimated errors, measured
within an aperture of radius r$_{eff}/2$, are presented in Table 2. The
galaxies marked with an asterisk are known to have emission lines, and
for these the values measured for the H$\beta$ indices are very uncertain,
and therefore not given.

\subsection{Emission lines}

Fig. 1 shows four of the five examples of emission-line elliptical
galaxies found in our sample. In each panel we show, at the top, the
wavelength calibrated spectrum, in the middle, a template spectrum
obtained by summing several stellar spectra and, at the bottom, the
subtraction of the two upper spectra, clearly showing the emission lines
H$\beta$ at 4861 {\rm \AA}  and [OIII] at 5007 {\rm \AA}.

Emission lines were visible in five of the 16 elliptical galaxies in
our sample. This may be compared to the study of 12 HCG elliptical
galaxies by Rubin, Hunter and Ford (1991). They found that emission
lines were present in 10 of those galaxies. It is tempting to compare
the frequency of emission lines found in these studies with that of
Phillips et al. (1986), who find a 60\% incidence of emission lines in
early-type galaxies in loose groups. However, a meaningful comparison
cannot be made between these results as the detection of faint emission
lines depends strongly on the resolution and S/N ratio of the spectrum.
For the same reason, our results are not inconsistent with those of
Rubin et al. (1991) since their spectra were  centered near 6500 \AA\,
allowing them to detect H$\alpha$, [NII], and S[II] emission lines.

  Emission lines were visible in 12 of the 28 (43\%) elliptical galaxies
from Gonz\'alez' sample. This fraction is similar to what we find for
the compact groups (5 in 16, 31\%) although the average S/N for
the Gonz\'alez' sample is almost an order of magnitude higher than ours.
The numbers of galaxies in the samples are still very small
and the data quality too inhomogeneous for a meaningful comparison
of the frequency of emission line elliptical galaxies in different
environments. 

\subsection{Mg$_b$, Fe5270 and Fe5335 indices}

   We plot in Fig. 2 the Mg$_b$ and the $<Fe>$ Lick/IDS indices
against velocity dispersion, for all galaxies in our sample and in
the control sample. In Fig. 3 is plotted the non-standard Lick/IDS
index defined by Gonz\'alez (1993), [MgFe], as a function of velocity
dispersion.  These plots show that the HCG elliptical galaxies have
the same mean Mg$_b$ and $<Fe>$ line strengths as do galaxies in the
control sample.  In both samples, the typical values indicate an excess
of Mg absorption with respect to the metallicity relation defined by the
galactic globulars.  This can be clearly seen in Fig. 4.  This figure
shows a plot of $Mg_b$ against $<Fe>$ line strengths for the compact
group and the control samples.  The continuous line is the locus of the
galactic globular clusters. The point at the bottom left of the diagram
is the galaxy HCG 04d.

     Two other studies have measured Mg and Fe line-strength
indices for galaxies in our compact group sample. Galaxies N7176 (or 90b)
and N7173 (or 90c) were measured by Burstein et al. (1987) and galaxies
14b, 86a and 86b were measured by Proctor et al. (2004).  A comparison
between our measurements and those other measurements, after employing
aperture corrections (Jorgensen et al. 1995) to transform all indices to
a common radius, are in agreement within 30$\%$,
or better.

\subsection{Simple stellar populations models with variable $[\alpha/Fe]$}

In normal luminous elliptical galaxies, the Mg excess has been
interpreted as an indication of a higher than solar [Mg/Fe] abundance
ratio, probably due to the dominant SNII over SNIa enrichment in
these galaxies (Gonz\'alez 1993; Worthey, Faber and Gonz\'alez 1992).
From the large dataset of nuclear line strengths of the Lick group,
presented in Trager et al. (1998), it is  shown that low velocity
dispersion objects exhibit solar [Mg/Fe] and more massive systems
are overabundant in magnesium.
 Luminous
HCG elliptical galaxies are not different in this respect.

Therefore, for the study of stellar populations in ellipticals it is 
crucial to employ population models that take into account $\alpha$-element
enhancements. 
Very few models that allow derivation of both [Z/H] and [$\alpha$/Fe],
or alternatively [Fe/H] and [$\alpha$/Fe], are available.
 
TMB03 present Simple Stellar
Population (SSP) models based
on the population synthesis code presented in Maraston (1998, 2005), in which
the fuel consumption theorem (Renzini \& Buzzoni 1986) is employed. The
models account for the $\alpha$-enhancement by using synthetic spectra
calculations by  Tripicco \& Bell (1995, hereafter TB95), and
 provide Lick indices for
SSPs covering ages from 1 to 15 Gyr, metallicities from 1/200 to 3.5
solar and  variable element abundance ratios.

 As an alternative approach, we computed a set of SSP models 
based on synthetic stellar spectra as described in
Barbuy et al. (2003). The computed  grid
of synthetic spectra covers the range 4600-5600 ${\rm \AA}$ for stellar
parameters 4000 $\leq$ T$_{\rm eff}$ $\leq$ 7000 K, 5.0 $\leq$ log g
$\leq$ 0.0, -3.0 $\leq$ [Fe/H] $\leq$ 0.3, and two values of chemical mixture 
[$\alpha$/Fe] = 0.0 and
+0.4 dex (the $\alpha$-elements considered are O,
Mg, Si, S, Ar, Ca, Ne, Ti). In the present work, a new
set of spectra with [$\alpha$/Fe] = +0.2 was computed,
 in order to better take into account the dependence of the indices on the
$\alpha$-enhancements [$\alpha$/Fe].  

The indices Mg$_2$, Fe5270 and Fe5335 were measured for the whole
parameter range and fitting functions of the form below 
(where $\theta$ = 5040/T$_{\rm eff}$) were derived
through a Levenberg-Marquadt algorithm

\smallskip
index = exp [$(a + b(log\theta) +
c(log\theta)^2 + d(log\theta)^3 + e(log g) +
f(log g)^2 + g(log g)^3
+ h([Fe/H]) + i([Fe/H]^2) + j([Fe/H]^3) + k([\alpha/Fe])
+ l(log \theta)(log g) + m(log \theta)([Fe/H])
+ n(log \theta)([\alpha/Fe])
+ o(log \theta)^2(log g) + p(log \theta)(log g)^2)$]
\smallskip

The new coefficients, 
which supersede the ones presented in Barbuy et al. (2003), are 
presented in Tables 3, 4 and 5.
The zero-point constants to calibrate the fitting functions to the 
Lick/IDS system were obtained following the  procedure described 
in detail in Barbuy et al. (2003, Section 3.1). Shortly, synthetic 
indices are computed with the fitting functions for the standard
Lick/IDS stars presented in Worthey et al. (1994). The input parameters
for the fitting funcions T$_{\rm eff}$, log g and [Fe/H]
are obtained directly from the catalog in Worthey et al. (1994).
The value of [$\alpha$/Fe] is adopted for each star as a function of [Fe/H],
given the $[Fe/H]$ vs. $[\alpha/Fe]$ relation presented in McWilliam (1997).
The synthetic indices computed are then compared with the measured
indices in the Lick/IDS system and a calibration constant can be obtained
by least-square fitting. The calibration constants are presented in Table 6.

Indices for SSPs were then derived by
combining these fitting functions with isochrones of Demarque et al.
(2004). The isochrones of Demarque et al. (2004) were preferred over
other $\alpha$-enhanced isochrones available in the literature (Salasnich
et al. 2000, Straniero et al. 1992, Salaris, Chieffi \&
Straniero 1993, Salaris \& Weiss 1998) because their chemical mixture
for the $\alpha$-enhancements are closer to the mixture adopted
in Barbuy et al. (2003) for the synthetic stellar grid. 
 Along the isochrone synthesis procedure, the
indices were weigthed by the
visual absolute magnitude (Greggio 1997).
The models were computed for a limited range of ages (4 $\leq$ t $\leq$ 16 Gyr)
and metallicities (0.008 $\leq$ Z $\leq$ 0.04), suitable to
the present analysis of the HCG galaxies. For this high metallicity
range, a strong contribution from blue horizontal
branch stars, not taken into account in our models, is not expected. 
  
One caveat to consider is that the fitting functions presented here, and
employed in the SSP models,
are strictly valid down to effective temperatures equal to 4000K, 
which is the lower limit of the temperature range of the synthetic grid 
in Barbuy et al. (2003). In cases where the isochrones require
 lower temperature stars, the fitting functions were extrapolated. 
A few
 stellar spectra for giants with temperatures of 3800K were computed,
employing Plez et al. (1992) model atmospheres. 
It was verified that the extrapolated stellar indices differ from the
measured ones typically by amounts of the order of 
the uncertainties on the fitting functions
(last line of Tables 3-5), and therefore the error
introduced in the integrated indices due to the extrapolation are
smaller than 10\%.
A further investigation of the behavior of these
indices in K and M stars will be discussed elsewhere. 
In particular, it is shown in Coelho et
al. (2005, in preparation) that TiO bands become strong in metal-rich
very cool stars, blending the Mg$_2$ feature, essentially preventing the use
of this index at very low stellar temperatures.

The SSP model indices for Mg$_2$,
Fe5270 and Fe5335 are given in Appendix A. 

As explained in detail in TMB03, their fitting functions are those by
Worthey et al. (1994), and corrections to account for the [$\alpha$/Fe]
enhancements were based on calculations of synthetic spectra by TB95 
(through the
so-called response functions),
following an extension of the method proposed by Trager et al. (1998).
The use of the fitting functions that are explicitly dependent on the
[$\alpha$/Fe] parameter 
is an alternative approach to employing empirical
fitting functions corrected by response functions (TMB03).  The former are
based on spectra that have all the $\alpha$-elements enhanced together,
while the latter incorporates changes in the indices due to individual
element variations.  It is not clear if the effect of the enhancement of
all $\alpha$-elements together is a linear combination of the effects
due to individual abundance variations. This question arises because an
important effect related to the $\alpha$-elements is that they are electron
donors. The continuum in G-M stars forms mainly by free-free and
bound-free transitions of H$^-$, whereas the amount of H$^-$ present
in their atmospheres depends on the electrons captured by Hydrogen atoms
which come from the $\alpha$-elements in particular (Fe is also an important
electron donor). Therefore, an overall enhancement of 
the $\alpha$-element abundance
can have an impact in the continuum (and therefore in the indices) that
is different from combining the impact of varying the abundance of each
element individually through response functions.  

\subsection{Stellar Populations of the HCG galaxies}

The indices measured from the spectra of sample galaxies
were compared to the SSP models,
and the luminosity weighted average ages $t$, metallicities [Z/H]
and $\alpha$-enhancements [$\alpha$/Fe] were derived.  

A fortran code
(kindly provided to us by D. Thomas) inspects correlations between
H$\beta$ and the metallic indices to find the best fit (t, [Z/H],
[$\alpha$/Fe]) for each galaxy. 

Only measurements with the smallest errors, i.e. H$\beta$ errors less
than 0.25 and fractional uncertainties for both Mg and Fe indices
lower than 10\%, were used in the derivation of the galaxy parameters.
Given the ages derived with the TMB03 models, values for [Fe/H], [Z/H]
and [$\alpha$/Fe] were also derived using the SSP models calculated in
this work, by selecting the best model through a $\chi$-square criteria
weighted by the measurements errors. The age estimations were based
entirely on TMB03 models since they carried out extensive comparisons
and corrections relative to observed models, which is needed for the use
of computed Hydrogen lines, and indices such as H$\beta$.  These derived
stellar population parameters are presented in Table 7 and the typical
estimated errors of the parameters are 2 Gyr for age and +0.2 dex for
[Z/H] and [$\alpha$/Fe].  The parameters derived with TMB03 models are
shown in columns 2 - 4 of Table 7 and those derived using our models are
shown in columns 5 - 7 of the same table.  For the galaxies with emission
and/or those whose errors on H$\beta$ were too large (three galaxies),
no age determination was attempted.

The problem of corrections of 
the galactic abundance pattern, as pointed out by Proctor et al. (2004)
in the use of SSP models by Vazdekis (1999), 
is not an issue in our case when the SSP models computed here
are employed, because this correction should be applied only to models that 
were built based on observed stars.
In addition, Thomas et al. (2005) concluded that corrections
for the local abundance pattern has no significant impact
on their SSP models, when [$\alpha$/Fe] is solar/supersolar.

There is a trend in the sense that the [$\alpha$/Fe] obtained by
the models calculated in this work are higher than those by TMB03 by 
$\approx$ 0.1 dex, and a comparison of the two sets of models employed 
can be seen in Fig. 5.
 Nevertheless, given the completely different ingredients of the two sets of
models (fitting functions, isochrones and SSP synthesis prescriptions),
the agreement is remarkable.

In Figs. 6 and 7 we plot the values for the HCG galaxies, the SSP models
from TMB03 and the corresponding values for the 
field galaxies from Gonz\'alez (1993) in the [MgFe]' vs. H${\beta}$  and
Mg$_b$ vs. $<Fe>$ diagrams respectively.

Though the statistics of the compact group galaxies is still small,
it is possible to see that the locus of the compact group galaxies
is characterized by old ages, solar and super-solar metallicities
and $\alpha$-enhancement. As a comparison, the field galaxies from the
sample of Gonz\'alez seem to span a wider  parameter range. However,
these also span a wider range of velocity dispersions (including less
massive galaxies), as can be seen in Figs. 2 and 3.

\section{Discussion}

\subsection{Possible selection effects} 

  One point that should be kept in mind is that the sample
of HCGs selected by Hickson (1982) is biased towards high surface
brightness {\it groups}.
At first sight, this may appear to preferentially
select the highest surface brightness galaxies, which, in turn, could
be biasing our results (to obtain preferentially old ages).

  In order to test this we have made a comparison of
the surface brightnesses within r$_{eff}$ 
of the galaxies in the control sample and
in the HCG sample. We found that  the two distributions can be driven
from the same parent population with a confidence level of 95\% 
(using the KS test), 
showing that 
we are comparing galaxies
of similar surface brightnesses, in the HCG and control samples. 

Another point that could be thought to bias our sample is the fact that
we discard galaxies which have emission lines when computing ages for
the objects.
This may select
against galaxies which have suffered minor accretion events, although
it is unlikely that we have missed major mergers (which would be
recognized also from their other optical properties).  This is
not a problem in Proctor et al's (2004) study, since they do not perform such
emission-line selection (because they use 20 different Lick indices for
determination of the galaxy parameters). The determination of old ages
for compact group galaxies seems, therefore, to be quite a robust result
and not a selection effect.

\subsection {Possible Outliers}

   The two best
candidate elliptical galaxies with young stellar populations and/or
mergers in our HCG sample are 04d and 14b, which are the galaxies with
the lowest velocity dispersions and lowest values of Mg$_b$.  Note that
in Figs. 2, 3 and 4, galaxies 04d, 14b are in all cases below
the line followed by the control sample galaxies. Galaxy 04d was studied
in detail by Zepf et al. (1991). Its surface brightness profile is well
fit by an r$^{1/4}$-law profile and it seems to be a genuine low-mass
elliptical galaxy (Zepf et al. 1991).  HCG 04d is one of the bluest
elliptical galaxies in Hickson's catalogue and it is significantly bluer
than galaxies of similar mass in other environments. Galaxy HCG 14b was
one of the few outliers in the diagram M$_8$ $\times$ r$_{eff}$ shown by
Mendes de Oliveira (1992), where M$_8$ is the absolute magnitude within
an aperture of radius equals to 8 h$^{-1}$ kpc, suggesting that this
galaxy has a very extended profile for its luminosity, resembling a cD
galaxy in this aspect.  This can be confirmed by the large measurement
of its effective radius.  However, Proctor et al. (2004) classify this
galaxy as a spiral bulge.  

\subsection{Other important work from the literature}

The $\alpha$-enhancement in galaxies, first pointed out by Peletier (1989)
and Worthey et al. (1992), was reviewed by Worthey (1998), Peletier
(1999) and Jorgensen (1999), from the observed Mg$_2$ vs. Fe indices.
Trager et al.  (1998), TMB03, and the present work derive
more quantitatively the $\alpha$-enhancements based on calculations of
synthetic spectra. The present sample galaxies also show overabundance
in magnesium-to-iron.

Several studies have tried to tackle the problem of environmental effects
on the spectral properties of elliptical galaxies.  Rose et al. (1994)
found that ellipticals/S0 in low density environments are considerably
more metal-rich, a result confirmed by Proctor et al.  (2004) and Thomas
et al. (2005). In addition, galaxies in the field have been found to be
younger than cluster counterparts (Bernardi et al.  1998; Proctor et
al. 2004; Thomas et al. 2005).  However, Proctor et
al. (2004) found that HCG galaxies are older than field galaxies and are more
consistent with the mean age of cluster galaxies.  In fact, our result
goes in the same direction: we find that the HCG galaxies are generally
older than their counterparts in looser environments.  

The fraction of possible ongoing mergers in compact groups has been
determined through  methods other than by studying the stellar
populations of the elliptical galaxies, as we are doing here.  The optical
colors of a sample of early-type galaxies and their infrared properties
were studied by Zepf et al. (1991) and Moles et al. (1994).  They found the
merging rate to be between 5 and 12\%. However, the candidate mergers they
present are most probably early-type galaxies with small contribution of
a young stellar component, instead of major mergers.  Our determinations
of ages, metallicities and $\alpha$-element abundances for compact group
galaxies also point to a low fraction of mergers in compact groups.
These results have important bearings on the age of the group, which is
discussed in the following.

\subsection {Implications of the results}

   Our result as well as Proctor et al.'s (2004) main result 
that HCG galaxies are generally old contradicts the expectations from
most formation scenarios for compact groups.  Diaferio et al. (1994)
proposed that loose groups may be continuously collapsing to form
compact groups, predicting that about 30\% of the elliptical galaxies in
compact groups should be mergers, which is not supported by the study of
the stellar populations of the elliptical galaxies presented here and
in Proctor et al. (2004).  One other scenario, proposed by Governato
et al. (1996), suggests that galaxies in compact groups merge to form
a massive elliptical galaxy and by secondary infall of surrounding
galaxies, the systems maintain their status as HCGs. According to this
scenario, merging of at least one or two massive galaxies happened at
redshifts between z=1 and 0.35, for their most probable realizations,
which would also be in disagreement with our results.  One scenario that
would be in agreement with our observations is that based on simulations
by Athanassoula et al. (1997), who suggest that if the dark matter is
distributed in a common group halo, rather than individual galaxy halos,
and assuming a high halo-to-total mass ratio and a density distribution
with very little central concentration, merging is suppressed and HCGs
would be dynamically very long-lived systems (i.e. they could live for
more than a Hubble time). Although the assumption of a common massive halo
around the group is supported by the observations, the lack of central
concentration is not (e.g. Zabludoff and Mulchaey 1998).
In fact, the opposite scenario is also supported by our findings, i.e.
one in which Hickson compact groups are young systems where merging
happens so fast (in a small fraction of the Hubble time) that few or no
merged objects are seen while the group is still recognizable as a group.

  One other important point to complement this discussion
is that compact groups exhibit a fraction of early-type galaxies
higher than that in the field or in loose groups (Hickson et al. 1988),
and more similar to clusters.  If galaxy-galaxy merging is not
responsible for the high early-type fractions, as suggested from our
results on the stellar populations, it is possible that the effects of the
environment are relatively unimportant at the current epoch and that the
similarity in the frequency of red galaxies and in the stellar populations
of galaxies in rich clusters and Hickson compact groups reflects 
conditions at the time of galaxy formation. 

\section{Acknowledgments} 
We are very grateful to Daniel Thomas and Robert Proctor for making
their codes available and for very helpful discussions.  We would also
like to thank Mike Bolte for useful comments on an early version of
the manuscript and an anonymous referee for suggested changes which
improved the paper. We thank Natarajan Visvanathan (in memoriam)
for obtaining the data for this study.  CMdO would like to acknowledge
support from the Brazilian FAPESP (projeto tem\'atico 01/07342-7), PRONEX,
the Alexander von Humboldt Foundation and the Max-Planck-Institut f\"ur
Extraterrestrische Physik, where this work was completed. PC acknowledges
a Fapesp PhD fellowship n$^{\circ}$ 2000/05237-9.  We made use of the
Hyperleda database and the NASA/IPAC Extragalactic Database (NED). The
latter is operated by the Jet Propulsion Laboratory, California Institute
of Technology, under contract with NASA.

%%%%%%%%%%%%%%%%%%%%%%%%%%%%%%%%%%%%%%%%%%%%%%%%%%%%%%%%%%%%%%%%%%%%%%%
%
%FIGURES
%
\clearpage
%%%%%%%%%%%%%%%%%%%%%%%%%%%%%%%%%%%%%%%%%%%%%%%%%%%%%%%%%%%%%%%%%
%%%%%%%%%%%%%%%%%%%%%%%%%%%%%%%%%%%%%%%%%%%%%%%%%%%%%%%%%%%%%%%%
%%
%%Figura 1
%%
%%%%%%%%%%%%%%%%%%%%%%%%%%%%%%%%%%%%%%%%%%%%%%%%%%%%%%%%%%%%%%%%%
%%%%%%%%%%%%%%%%%%%%%%%%%%%%%%%%%%%%%%%%%%%%%%%%%%%%%%%%%%%%%%%%

\begin{figure*}
\plotone{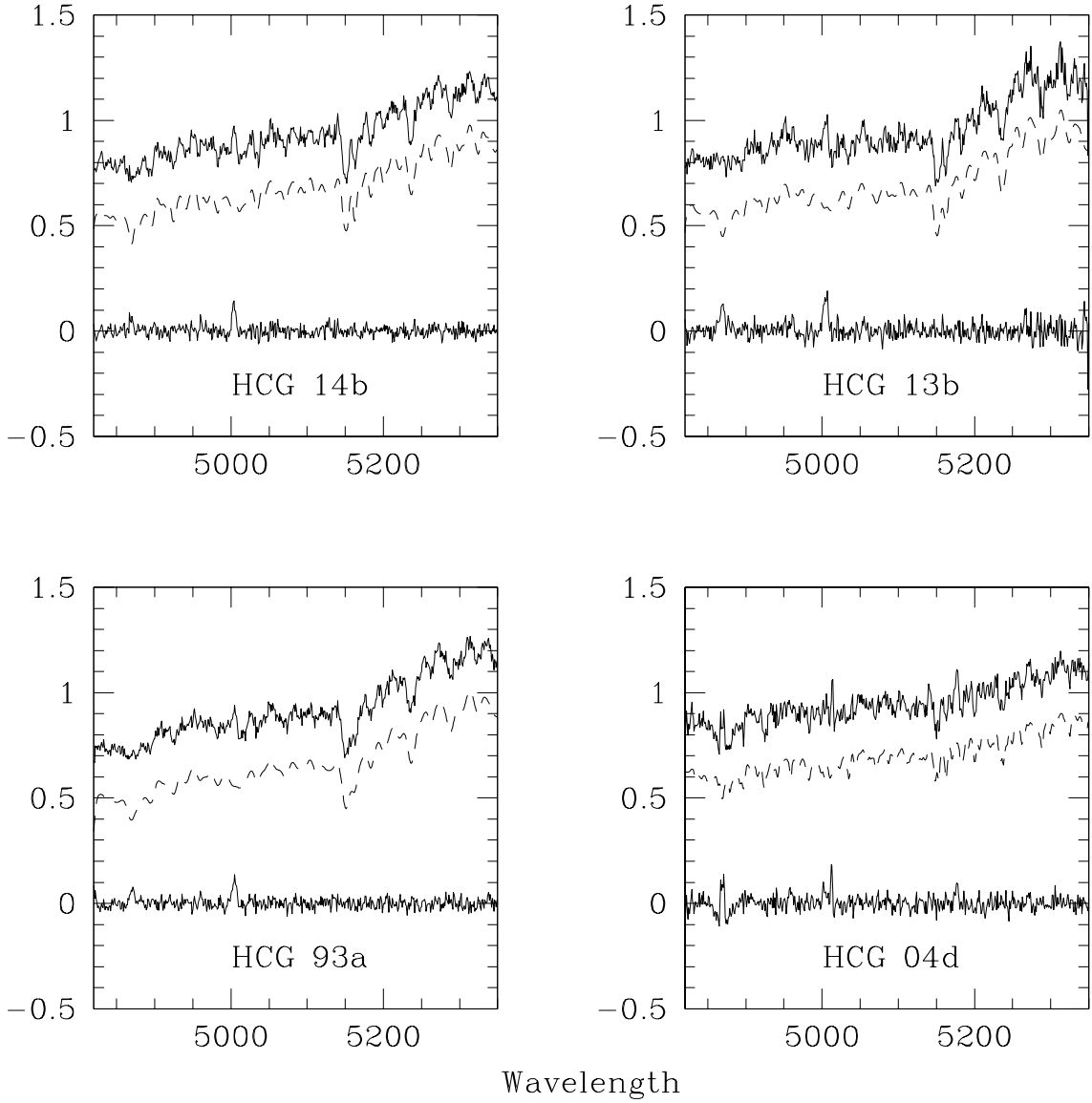}
\figcaption{
Spectra of four selected galaxies in our sample, which
present emission lines.
For each panel: top line -- wavelength calibrated spectrum; middle (in broken lines) --
optimal template; bottom line --
difference between the spectrum and the template.
In the latter one can easily
see the lines in emission (H$\beta$ at 4861 $\rm{\AA}$ and [OIII]
at 5007 $\rm{\AA}$)
Five of the 16 elliptical galaxies in our sample
showed emission lines in their spectra. The y axis is an arbitrary
scale.
}
 \end{figure*}

%%%%%%%%%%%%%%%%%%%%%%%%%%%%%%%%%%%%%%%%%%%%%%%%%%%%%%%%%%%%%%%%
%%
%%Figura 2
%%
%%%%%%%%%%%%%%%%%%%%%%%%%%%%;%%%%%%%%%%%%%%%%%%%%%%%%%%%%%%%%%%%%%
%%%%%%%%%%%%%%%%%%%%%%%%%%%%%%%%%%%%%%%%%%%%%%%%%%%%%%%%%%%%%%%%

\begin{figure*}
\plotone{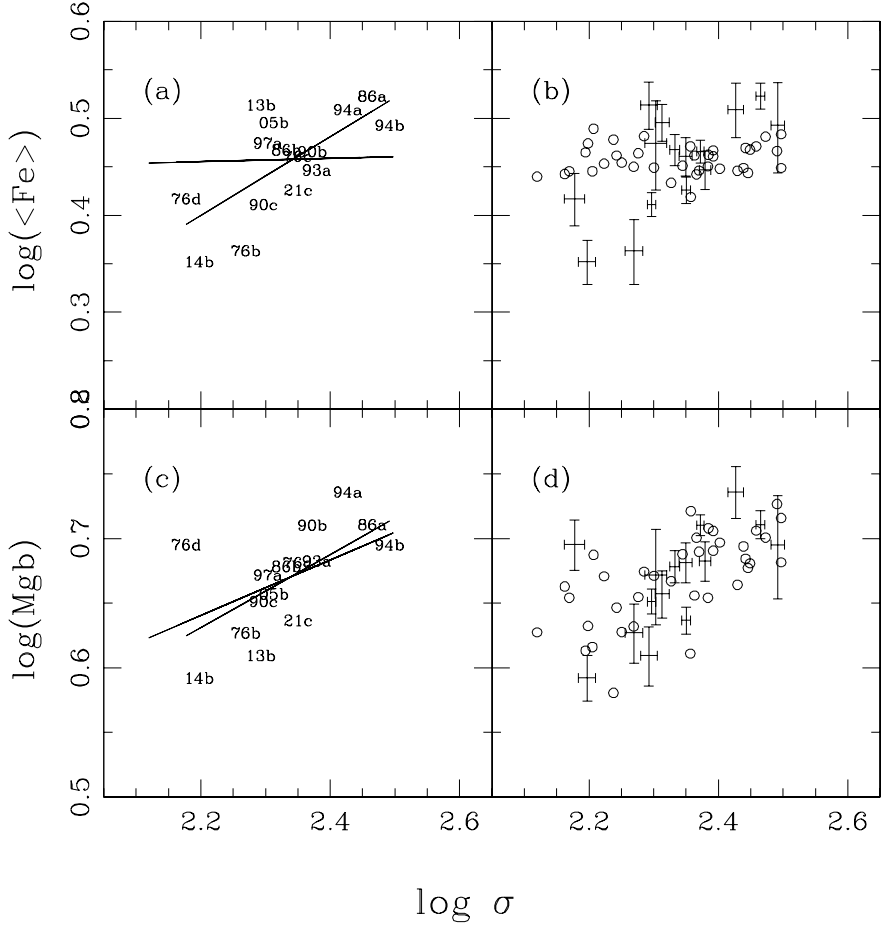}
\figcaption{
A plot of $<Fe>$ and $Mg_b$ $vs.$ velocity dispersion for 16 galaxies in compact
groups (galaxy 04d is not plotted; its indices are out of the scale of
the plot). Diagrams (a) and (c) identify the objects by their Hickson
catalogue ID numbers.  The straight lines are the mean relations for the
compact group and the control samples. In diagrams (b) and (d) the open
circles indicate the galaxies in the control sample and the center of the
error bars indicate the galaxies in the compact groups.
}
\end{figure*}

%%%%%%%%%%%%%%%%%%%%%%%%%%%%%%%%%%%%%%%%%%%%%%%%%%%%%%%%%%%%%%%%
%%
%%Figura 3 
%%
%%%%%%%%%%%%%%%%%%%%%%%%%%%%%%%%%%%%%%%%%%%%%%%%%%%%%%%%%%%%%%%%%
%%%%%%%%%%%%%%%%%%%%%%%%%%%%%%%%%%%%%%%%%%%%%%%%%%%%%%%%%%%%%%%%

\begin{figure*}
\plotone{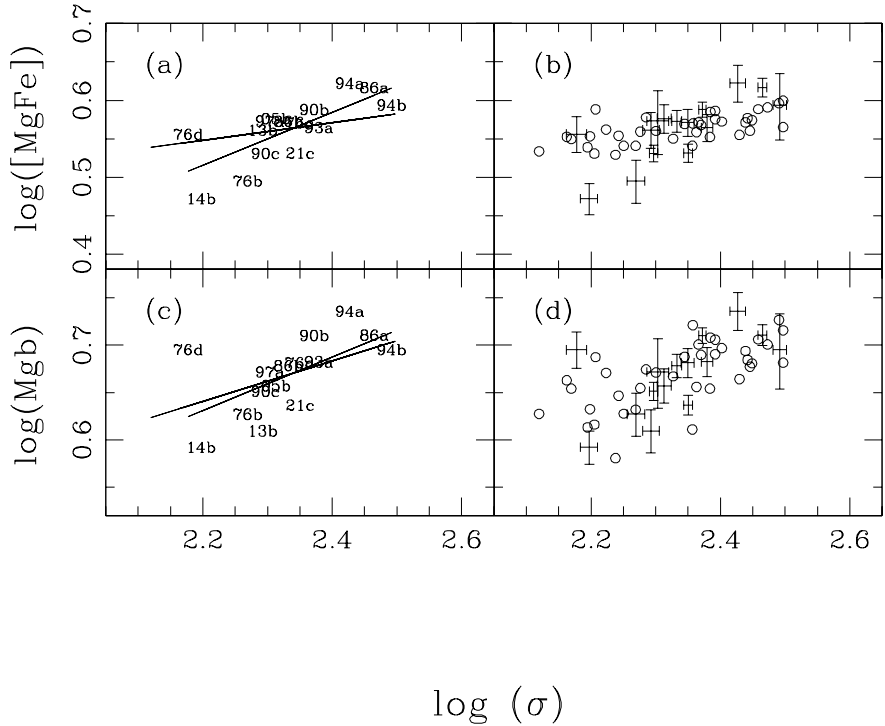}
\figcaption[f2b.ps]{
A plot of [MgFe] and $Mg_b$ $vs.$ velocity dispersion for 16 galaxies in compact
groups (galaxy 04d is not plotted; its indices are out of the scale of
the plot).
 The index [MgFe] is defined as [MgFe] = sqrt(Mg b . $<Fe>$)
by Gonz\'alez (1993).
 Diagrams (a) and (c) identify the objects by their Hickson
catalogue ID numbers.  The straight lines are the mean relations for the
compact group and the control samples. In diagrams (b) and (d) the open
circles indicate the galaxies in the control sample and the center of the
error bars indicate the galaxies in the compact groups.
}
\end{figure*}

%%%%%%%%%%%%%%%%%%%%%%%%%%%%%%%%%%%%%%%%%%%%%%%%%%%%%%%%%%%%%%%%
%%
%%Figura 4 
%%
%%%%%%%%%%%%%%%%%%%%%%%%%%%%%%%%%%%%%%%%%%%%%%%%%%%%%%%%%%%%%%%%%
%%%%%%%%%%%%%%%%%%%%%%%%%%%%%%%%%%%%%%%%%%%%%%%%%%%%%%%%%%%%%%%%

\begin{figure*}
\plotone{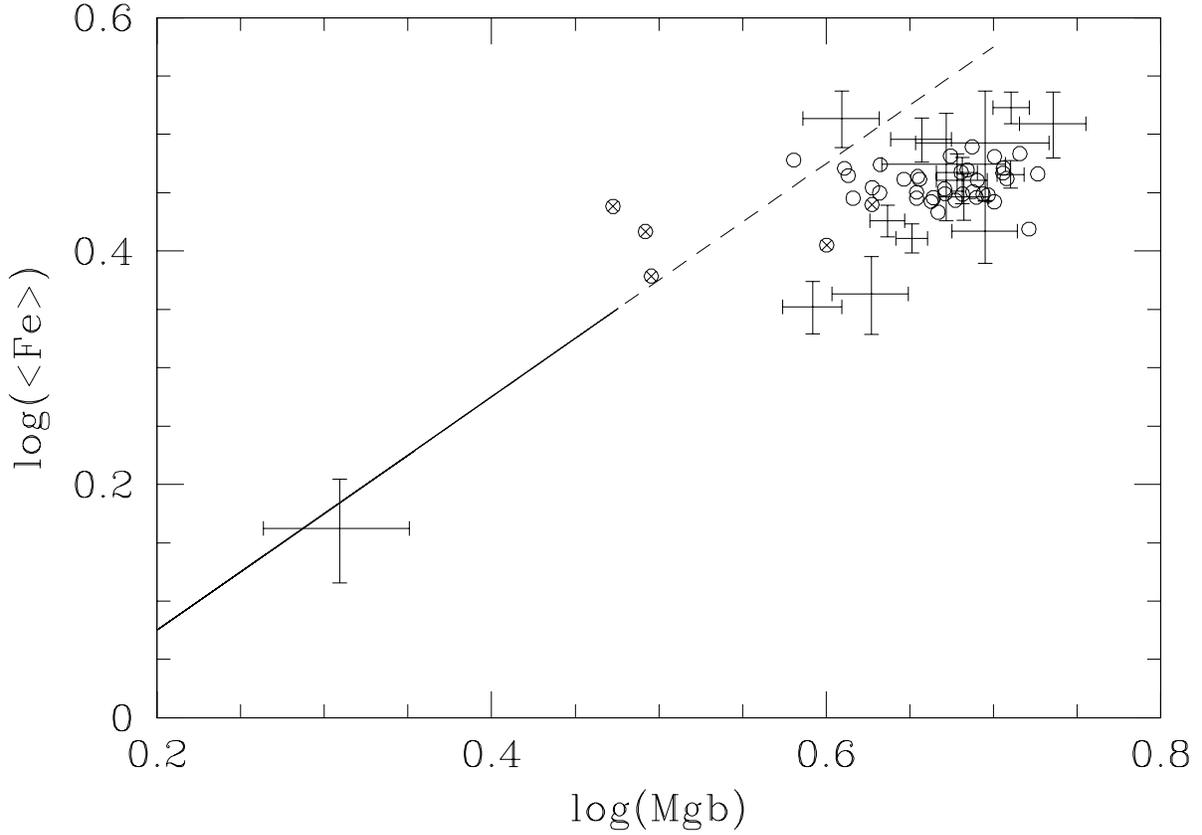}
\figcaption[f3.ps]{
Plot of $Mg_b$
against $<Fe>$ line strengths for the compact groups (indicated by the
error bars) and the
control sample (open circles). Four galaxies from the control sample
with velocity dispersions smaller than 100 km/s are indicated as
crossed circles (these are not included in the previous figures). The
continuous line is the locus of the galactic globular clusters
and the broken line is an extrapolation of it.
}
 \end{figure*}

%%%%%%%%%%%%%%%%%%%%%%%%%%%%%%%%%%%%%%%%%%%%%%%%%%%%%%%%%%%%%%%%
%%
%%Figura 5 
%%
%%%%%%%%%%%%%%%%%%%%%%%%%%%%%%%%%%%%%%%%%%%%%%%%%%%%%%%%%%%%%%%%%
%%%%%%%%%%%%%%%%%%%%%%%%%%%%%%%%%%%%%%%%%%%%%%%%%%%%%%%%%%%%%%%%

\begin{figure*}
\plotone{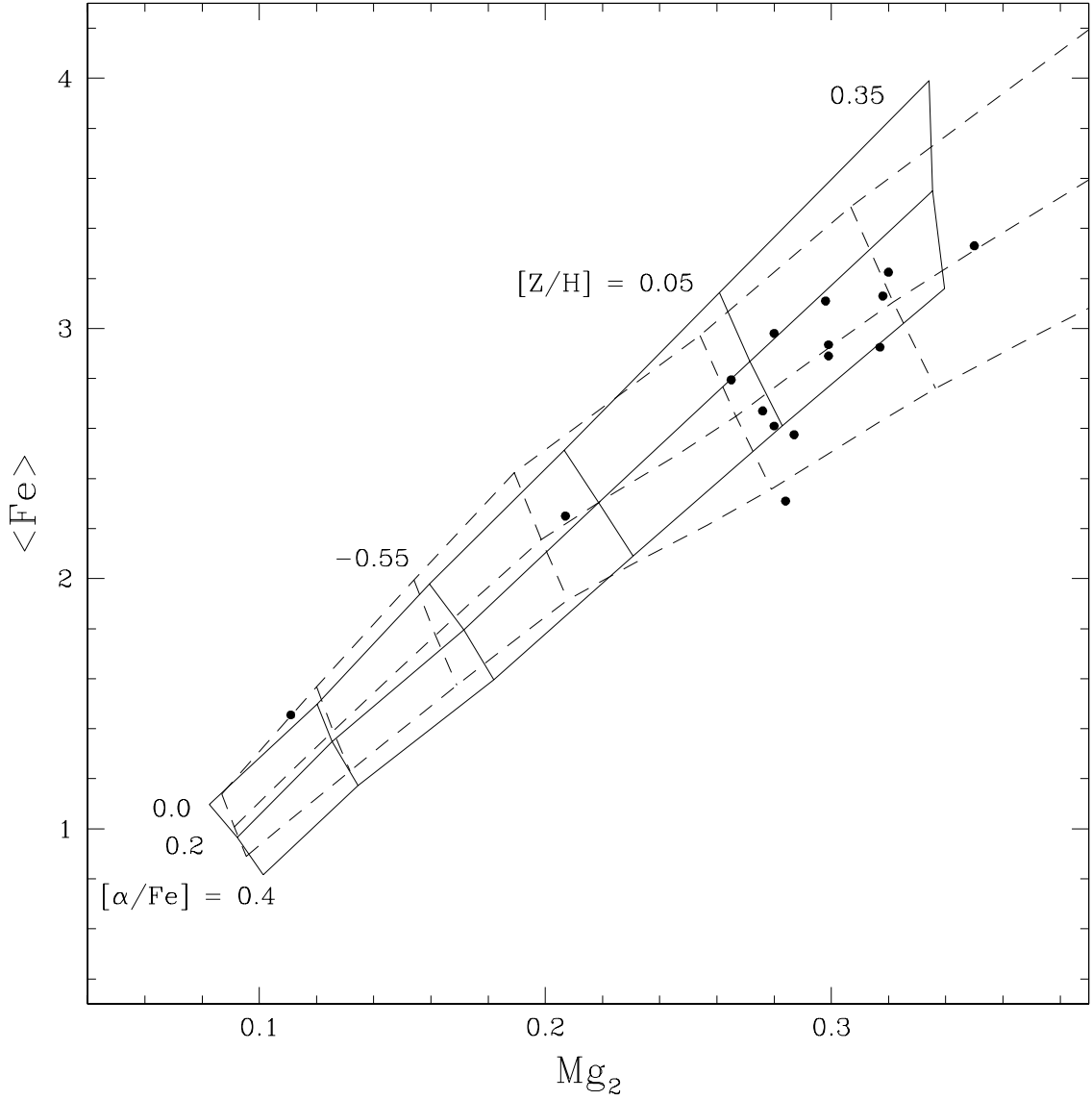}
\figcaption[f4.ps]{
Compact group galaxies (filled circles) in a $<$Fe$>$ vs. $Mg_2$ diagram,
overplotted with two sets of
 SSP models, where the TMB03 models are
the dashed lines, and models computed in this work are the solid lines.
Labels are given to illustrate the loci occupied by
different stellar population parameters.
The models are plotted for an age t = 12 Gyr.
}
\end{figure*}

%%%%%%%%%%%%%%%%%%%%%%%%%%%%%%%%%%%%%%%%%%%%%%%%%%%%%%%%%%%%%%%%
%%
%%Figura 6 
%%
%%%%%%%%%%%%%%%%%%%%%%%%%%%%%%%%%%%%%%%%%%%%%%%%%%%%%%%%%%%%%%%%%
%%%%%%%%%%%%%%%%%%%%%%%%%%%%%%%%%%%%%%%%%%%%%%%%%%%%%%%%%%%%%%%%

\begin{figure*}
\plotone{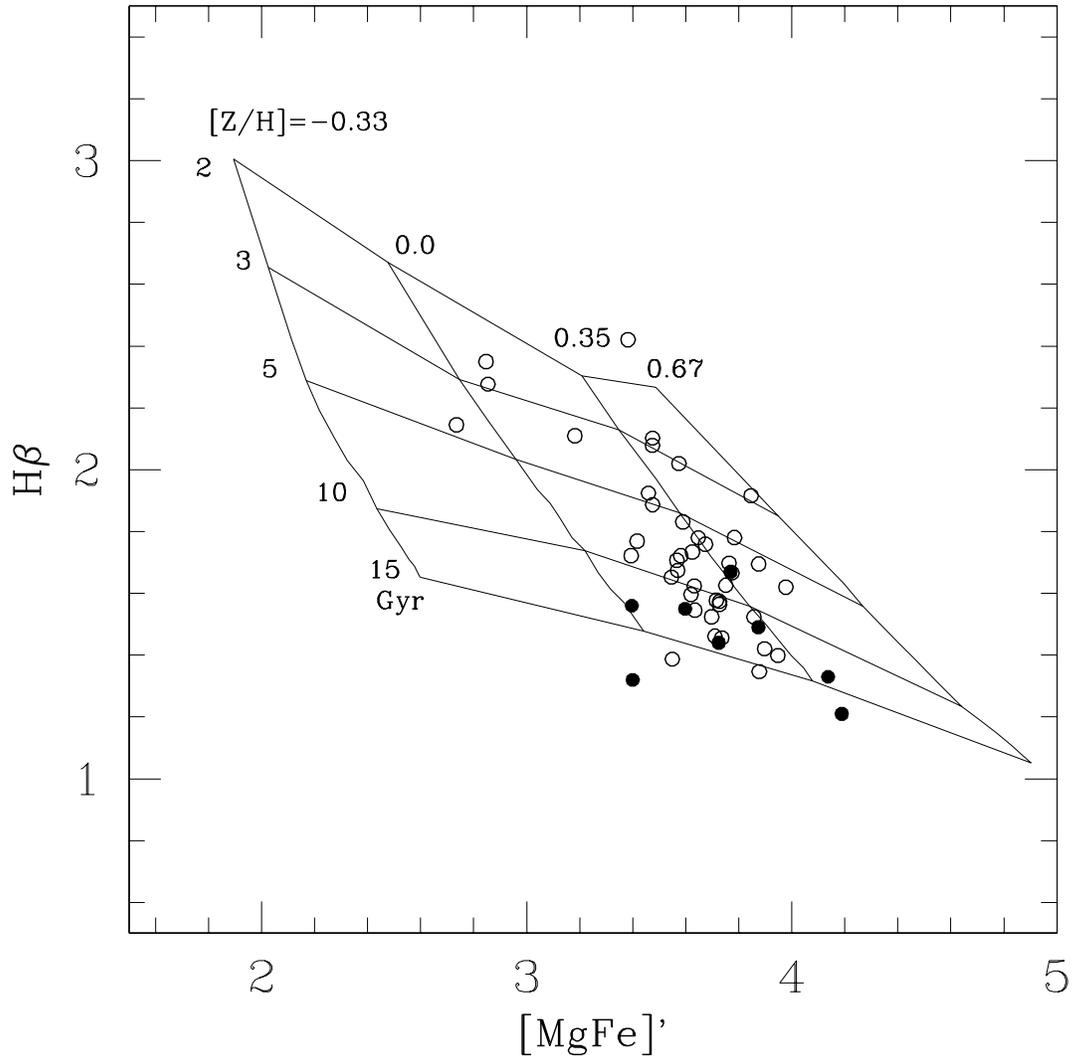}
\figcaption[f5.ps]{
Compact group galaxies (filled circles), TMB03
models (grid)
and field galaxies (open circles) in H$\beta$ vs. [MgFe]' diagram.
 The index [MgFe]'
follows TMB03 and is defined as [MgFe]' = (Mg$b$(0.72 Fe5270 +
 0.28 Fe5335). Labels are given to illustrate the loci occupied by
different stellar population parameters.
 In each model sequence, ages increase from top-left
to bottom-right.
}
\end{figure*}

%%%%%%%%%%%%%%%%%%%%%%%%%%%%%%%%%%%%%%%%%%%%%%%%%%%%%%%%%%%%%%%%
%%
%%Figura 7 
%%
%%%%%%%%%%%%%%%%%%%%%%%%%%%%%%%%%%%%%%%%%%%%%%%%%%%%%%%%%%%%%%%%%
%%%%%%%%%%%%%%%%%%%%%%%%%%%%%%%%%%%%%%%%%%%%%%%%%%%%%%%%%%%%%%%%

\begin{figure*}
\plotone{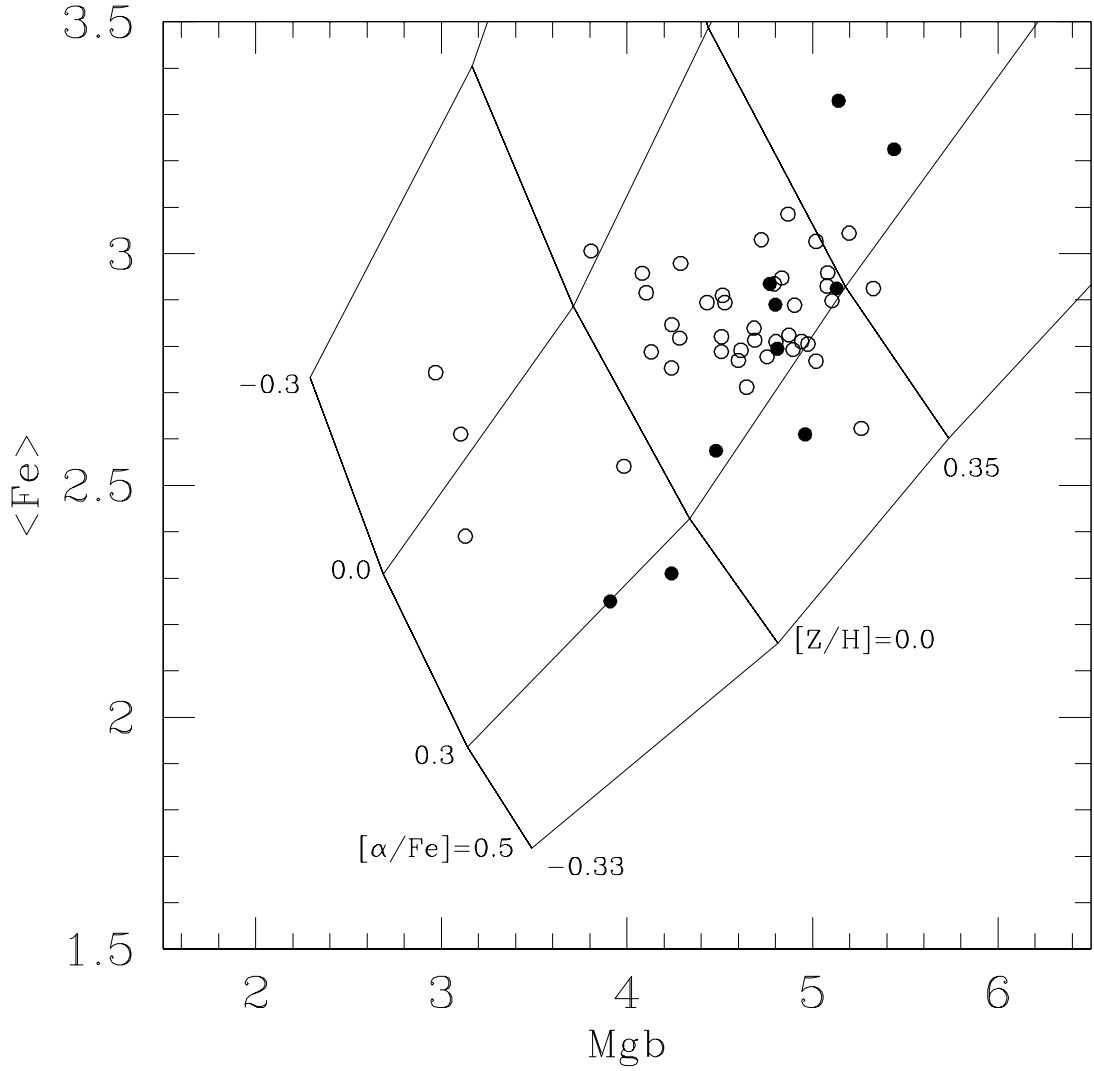}
\figcaption[f6.ps]{
Compact group galaxies (filled circles), TMB03 models (grid)
and field galaxies (open circles) in a Mg$b$ vs. $<$Fe$>$ diagram.
 Labels are given to illustrate the loci occupied by
different stellar population parameters.
The models are plotted for an age t = 12 Gyr.
}
\end{figure*}

\clearpage

%%%%%%%%%%%%%%%%%%%%%%%%%%%%%%%%%%%%%%%%%%%%%%%%%%%%%%%%%%%%%%%%%%%%
%%%%%%%%%%%%%%%%%%%%%%%%%%%%%%%%%%%%%%%%%%%%%%%%%%%%%%%%%%%%%%%%%%%%
%
%Table 1 Main parameters for each galaxy
%
%%%%%%%%%%%%%%%%%%%%%%%%%%%%%%%%%%%%%%%%%%%%%%%%%%%%%%%%%%%%%%%%%%%%
%%%%%%%%%%%%%%%%%%%%%%%%%%%%%%%%%%%%%%%%%%%%%%%%%%%%%%%%%%%%%%%%%%%%

\begin{deluxetable}{lcccccc}                                                    %\footnotesize
\small
\tablenum{1} \tablewidth{0pt}
\tablecaption{Main photometric and kinematic parameters for each galaxy}
\tablecolumns{7}
\tablehead{                                                                     \colhead{ID} &
\colhead{$r_{eff}$} &                                                           \colhead{$\mu_{eff}$} &
\colhead{S/N} &                                                                 \colhead{V$_{rad}$} &
\colhead{$\sigma$} &                                                            \colhead{$\Delta\sigma$}
\\                                                                              \colhead{} &
\colhead{in arcsec} &                                                           \colhead{Type (Hickson)} &
\colhead{Type (NED)}  &                                                         \colhead{} &
\colhead{$km~s^{-1}$} &                                                         \colhead{$km~s^{-1}$} }
                                                                                \startdata
%04C & \nodata & \nodata & 27 & 17789.8 &  174.8 & 14.8 \\
04D* & 4.0 & 20.22 & 30 & 8005.7 &  71.8  & 5.2 \\
05B & 7.0 & 21.21 & 49 & 11944.7 &  205.6  & 5.2 \\
13B* & 8.5 & 21.83 & 36 & 12114.2 &  196.1 & 5.8 \\
14B* & 64.0 & 24.06 & 38 & 5188.8 &  157.5 & 4.8 \\
21C & 21.0 & 22.83 & 77 & 7208.3 &  224.0 & 3.4 \\
76B & 10.0 & 21.47 & 36 & 9882.1 &  186.0 & 5.8 \\
76C & 12.0 & 22.07 & 45 & 10473.8 &  223.5 & 5.2 \\
76D &  6.0 & 21.52 & 30 & 10108.9 &  150.5  & 5.4 \\
86A & 15.7 & 21.33 & 70 & 5923.1 &  291.7  & 4.4 \\
86B* & 8.0 & 20.70 & 54 & 5805.2 &  215.0  & 3.7 \\
90B & 12.0 & 20.40 & 75 & 2502.0 &  235.6 & 3.2  \\
90C & 14.7 & 20.73 & 74 & 2550.9 &  198.0  & 2.9 \\
93A* & 22.0 & 21.78 & 54 & 4956.7 &  239.5  & 4.9 \\
94A & 11.7 & 21.91 & 39 & 11765.1 &  267.3  & 7.3 \\
94B & 11.9 & 21.86 & 42 & 11783.8 &  310.2  & 7.3 \\
97A & 40.2 & 23.27 & 23 & 6818.3 &  201.1  & 7.8  \\
\enddata

\tablenotetext{(*)}{ 
Has emission lines
}

\end{deluxetable}

%%%%%%%%%%%%%%%%%%%%%%%%%%%%%%%%%%%%%%%%%%%%%%%%%%%%%%%%%%%%%%%%%%%%
%%%%%%%%%%%%%%%%%%%%%%%%%%%%%%%%%%%%%%%%%%%%%%%%%%%%%%%%%%%%%%%%%%%%
%
%Table 2 Line indices
%
%%%%%%%%%%%%%%%%%%%%%%%%%%%%%%%%%%%%%%%%%%%%%%%%%%%%%%%%%%%%%%%%%%%%
%%%%%%%%%%%%%%%%%%%%%%%%%%%%%%%%%%%%%%%%%%%%%%%%%%%%%%%%%%%%%%%%%%%%

\begin{deluxetable}{rccccccrr}
\tablenum{2} \tablewidth{0pt}
\tablecaption{Line indices for the compact group galaxies}
\tablecolumns{9}
\tablehead{
\colhead{ID} &
\colhead{$H\beta$} &
\colhead{$Mg_1$} &
\colhead{$Mg_2$} &
\colhead{$Mg_b$} &
\colhead{$Fe_{5015}$} &
\colhead{$Fe_{5270}$} &
\colhead{$Fe_{5335}$} &
\colhead{$Fe_{5406}$}
 }
\startdata
04D     & \nodata  &  0.055$\pm$0.005 &  0.111$\pm$0.005 &  2.04$\pm$0.20 &    3.15$\pm$0.41 &  1.32$\pm$0.23 &  1.59$\pm$0.19 &  0.90$\pm$0.14 \\
05B     &  1.67$\pm$0.23 &  0.156$\pm$0.004 &  0.318$\pm$0.005 &  4.54$\pm$0.19 &    4.94$\pm$0.40 &  3.36$\pm$0.19 &  2.90$\pm$0.20 &  1.83$\pm$0.15 \\
13B     & \nodata &  0.149$\pm$0.005 &  0.282$\pm$0.006 &  4.07$\pm$0.21 &   4.05$\pm$0.42 &  3.39$\pm$0.22 &  3.14$\pm$0.28 &  1.74$\pm$0.26 \\
14B     &  \nodata  &  0.083$\pm$0.004 &  0.207$\pm$0.004 &  3.91$\pm$0.16 &    4.08$\pm$0.34 &  2.39$\pm$0.17 &  2.11$\pm$0.16 &  1.37$\pm$0.12 \\
21C     &  1.32$\pm$0.12 &  0.121$\pm$0.002 &  0.276$\pm$0.003 &  4.33$\pm$0.10 &    4.93$\pm$0.24 &  2.92$\pm$0.11 &  2.42$\pm$0.12 &  1.75$\pm$0.10 \\
76B     &  1.23$\pm$0.29 &  0.133$\pm$0.005 &  0.284$\pm$0.006 &  4.24$\pm$0.22 &    5.64$\pm$0.45 &  2.58$\pm$0.22 &  2.04$\pm$0.27 &  1.57$\pm$0.22 \\
76C     &  1.44$\pm$0.16 &  0.144$\pm$0.004 &  0.299$\pm$0.004 &  4.80$\pm$0.17 &    5.88$\pm$0.36 &  2.82$\pm$0.17 &  2.96$\pm$0.20 &  1.92$\pm$0.15 \\
76D     &  1.55$\pm$0.22 &  0.124$\pm$0.005 &  0.280$\pm$0.006 &  4.96$\pm$0.22 &    3.81$\pm$0.53 &  2.61$\pm$0.23 &  1.32$\pm$0.23 &  1.32$\pm$0.27 \\
86A     &  1.33$\pm$0.12 &  0.172$\pm$0.002 &  0.350$\pm$0.003 &  5.14$\pm$0.13 &    6.09$\pm$0.28 &  3.60$\pm$0.13 &  3.06$\pm$0.16 &  1.81$\pm$0.13 \\
86B     &  \nodata &  0.136$\pm$0.003 &  0.299$\pm$0.004 &  4.77$\pm$0.14 &    4.79$\pm$0.30 &  2.95$\pm$0.15 &  2.92$\pm$0.16 &  1.69$\pm$0.13 \\
90B     &  1.49$\pm$0.12 &  0.158$\pm$0.002 &  0.317$\pm$0.003 &  5.13$\pm$0.10 &    5.67$\pm$0.21 &  3.17$\pm$0.10 &  2.68$\pm$0.12 &  1.92$\pm$0.10 \\
90C     &  1.56$\pm$0.11 &  0.131$\pm$0.002 &  0.287$\pm$0.003 &  4.48$\pm$0.10 &    4.88$\pm$0.22 &  2.81$\pm$0.10 &  2.34$\pm$0.11 &  1.84$\pm$0.08 \\
93A     &  \nodata  &  0.113$\pm$0.003 &  0.265$\pm$0.004 &  4.81$\pm$0.17 &    4.44$\pm$0.36 &  3.04$\pm$0.18 &  2.55$\pm$0.18 &  1.91$\pm$0.13 \\
94A     &  1.21$\pm$0.22 &  0.159$\pm$0.005 &  0.320$\pm$0.007 &  5.44$\pm$0.25 &    5.28$\pm$0.59 &  3.54$\pm$0.26 &  2.91$\pm$0.33 &  1.67$\pm$0.26 \\
94B     &  0.57$\pm$0.32 &  0.168$\pm$0.009 &  0.298$\pm$0.012 &  4.96$\pm$0.45 &    5.82$\pm$1.00 &  3.40$\pm$0.46 &  2.82$\pm$0.49 &  1.68$\pm$0.35 \\
97A     &  1.92$\pm$0.43 &  0.124$\pm$0.008 &  0.280$\pm$0.010 &  4.70$\pm$0.40 &    4.74$\pm$0.79 &  3.28$\pm$0.40 &  2.68$\pm$0.49 &  2.01$\pm$0.29 \\
\enddata
\end{deluxetable}

%%%%%%%%%%%%%%%%%%%%%%%%%%%%%%%%%%%%%%%%%%%%%%%%%%%%%%%%%%%%%%%%%%%%
%%%%%%%%%%%%%%%%%%%%%%%%%%%%%%%%%%%%%%%%%%%%%%%%%%%%%%%%%%%%%%%%%%%%
%
%Table 3-6 Fitting functions
%
%%%%%%%%%%%%%%%%%%%%%%%%%%%%%%%%%%%%%%%%%%%%%%%%%%%%%%%%%%%%%%%%%%%%
%%%%%%%%%%%%%%%%%%%%%%%%%%%%%%%%%%%%%%%%%%%%%%%%%%%%%%%%%%%%%%%%%%%%

\clearpage
\begin{deluxetable}{llcccc}
\tablenum{3} \tablewidth{0pt}
\tablecaption{Fitting function coefficients for Mg$_2$ index.}
\label{ff1}
\tablehead{
&Validity	&5000 $\leq$ $T_{\rm eff}$ $\leq$ 7000	&5000 $\leq$ T$_{\rm eff}$ $\leq$ 7000	&4000 $\leq$ T$_{\rm eff}$ $\leq$ 5000(*)  \\
&range		& 0.0 $\leq$ log $g$ $\leq$ 5.0		& 0.0 $\leq$ log $g$ $\leq$ 5.0 	& 0.0 $<$ log $g$ $\leq$ 5.0		\\
&		& -3.0 $\leq$ [Fe/H] $\leq$ -1.0	& -1.0 $\leq$ [Fe/H] $\leq$ +0.3	& -3.0 $\leq$ [Fe/H] $\leq$ +0.3        \\
Coefficient & Term &  &   & }
\startdata
a & constant                      	& -2.4823	& -2.5161 	& 0.0141 	\\
b & log $\theta$                   	& -2.7275 	&  0.0104 	&-2.1352 	\\
c & (log $\theta$)$^2$              	& 0        	& 22.2153 	&37.7958 	\\
d & (log $\theta$)$^3$              	& 0       	& 0       	& 0      	\\
e & log $g$                       	& -0.0218 	&-0.1128 	&-0.0616 	\\
f & (log $g$)$^2$                  	& 0      	& 0.0421	& 0.0185  	\\
g & (log $g$)$^3$                  	& 0.01043	& 0.0088	& 0       	\\
h &  ${\rm [Fe/H]}$               	& 0.41801	& 0.6361 	& 0.0899 	\\
i & ${\rm [Fe/H]}^2$             	&-0.112  	& 0		& 0      	\\
j & ${\rm [Fe/H]}^3$             	& 0      	& 0		& 0      	\\
k & ${\rm [\alpha /Fe]}$          	& 0.8161 	& 0.81051	& 0.0904 	\\
l & (log $\theta$)(log $g$)        	& 4.5051 	& 3.0778    	& 2.4707 	\\
m & (log $\theta$)(${\rm [Fe/H]}$) 	& 1.4022 	& 1.4763 	& 0.7067 	\\
n & (log $\theta$)([$\alpha$/Fe])  	& 4.3325 	& 3.8713 	& 1.4655 	\\
o & (log $\theta$)$^2$(log $g$)     	& 14.560  	& 6.6195  	&-11.4960 	\\
p & (log $\theta$)(log $g$)$^2$     	& 0.1697 	& 0.2266 	&-0.2505 	\\
\hline
 & r.m.s. & 0.007 	&  0.014	& 0.038 	\\
\enddata
\tablenotetext{(*)}{ 
This interval is not described in terms of exponential. The index is given directly by the polynomial form inside the parenthesis in the equation in the text.
}
\end{deluxetable}

\begin{deluxetable}{llcccc}
\tablenum{4} \tablewidth{0pt}
\tablecaption{Fitting function coefficients for Fe5270 index.}
\label{ff2}
\tablehead{
Validity	&&4000 $\leq$ $T_{\rm eff}$ $\leq$ 7000	&4000 $\leq$ T$_{\rm eff}$ $\leq$ 4750	&4750 $\leq$ T$_{\rm eff}$ $\leq$ 7000        \\
range	&	& 0.0 $\leq$ log $g$ $\leq$ 3.0		& 3.0 $\leq$ log $g$ $\leq$ 5.0 	& 3.0 $<$ log $g$ $\leq$ 5.0		\\
&		& -3.0 $\leq$ [Fe/H] $\leq$ +0.3	& -3.0 $\leq$ [Fe/H] $\leq$ +0.3	& -3.0 $\leq$ [Fe/H] $\leq$ +0.3        \\
Coefficient & Term &  &   & 
}
\startdata
a & constant                      	& 1.6686 	&  0.4078 	& 1.4406 	\\
b & log $\theta$                   	& -0.6681 	& 10.8621 	&-1.3135 	\\
c & (log $\theta$)$^2$              	& 7.2196       	&-18.6671 	&36.3395 	\\
d & (log $\theta$)$^3$              	& 56.9337      	& 0       	& 0      	\\
e & log $g$                       	& -0.2003 	& 0.1996 	&-0.2830 	\\
f & (log $g$)$^2$                  	& 0      	& 0     	& 0.0509  	\\
g & (log $g$)$^3$                  	& 0      	& 0             & 0       	\\
h &  ${\rm [Fe/H]}$               	& 0.5467 	& 0.5234 	& 0.46607	\\
i & ${\rm [Fe/H]}^2$             	&-0.0316 	& 0		&-0.0150 	\\
j & ${\rm [Fe/H]}^3$             	& 0.0163 	& 0.0284        & 0.0310      	\\
k & ${\rm [\alpha /Fe]}$          	& 0.0493 	&-0.1458 	&-0.0755 	\\
l & (log $\theta$)(log $g$)        	& 0.5784 	&-1.2399    	& 1.9521 	\\
m & (log $\theta$)(${\rm [Fe/H]}$) 	&-1.7989 	&-2.7461 	&-2.6080 	\\
n & (log $\theta$)([$\alpha$/Fe])  	& 0      	&-1.3704 	&-1.6855 	\\
o & (log $\theta$)$^2$(log $g$)     	& 4.3268 	& 0       	&-3.4429 	\\
p & (log $\theta$)(log $g$)$^2$     	& 0.2522 	& 0      	& 0      	\\
\hline
 & r.m.s. & 0.20 	& 0.22 	& 0.07 	\\
\enddata
\end{deluxetable}

\begin{deluxetable}{llcccc}
\tablenum{5} \tablewidth{0pt}
\tablecaption{Fitting function coefficients for Fe5335 index.}
\label{ff3} 
\tablehead{
Validity	& &4000 $\leq$ $T_{\rm eff}$ $\leq$ 7000	&4000 $\leq$ T$_{\rm eff}$ $\leq$ 4750	&4750 $\leq$ T$_{\rm eff}$ $\leq$ 7000        \\
range	&	& 0.0 $\leq$ log $g$ $\leq$ 3.0		& 3.0 $\leq$ log $g$ $\leq$ 5.0 	& 3.0 $<$ log $g$ $\leq$ 5.0		\\
&		& -3.0 $\leq$ [Fe/H] $\leq$ +0.3	& -3.0 $\leq$ [Fe/H] $\leq$ +0.3	& -3.0 $\leq$ [Fe/H] $\leq$ +0.3        \\
Coefficient & Term &  &   & 
}
\startdata
a & constant                      	& 1.5212 	&  0.2631 	& 1.2824 	\\
b & log $\theta$                   	& 1.2643 	& 13.5966 	& 0      	\\
c & (log $\theta)^2$              	&10.4707       	&-23.3170 	&38.6002 	\\
d & (log $\theta)^3$              	&38.8133       	& 0       	& 0      	\\
e & log $g$                       	& -0.1578 	& 0.24554	&-0.2363 	\\
f & (log $g$)$^2$                 	&-0.0077 	& 0     	& 0.0492  	\\
g & (log $g$)$^3$                  	& 0      	& 0             & 0       	\\
h &  ${\rm [Fe/H]}$               	& 0.5273 	& 0.5849 	& 0.4698 	\\
i & ${\rm [Fe/H]}^2$             	& 0      	&-0.07951     	&-0.0202 	\\
j & ${\rm [Fe/H]}^3$             	& 0.0304 	& 0             & 0.0333      	\\
k & ${\rm [\alpha /Fe]}$          	&-0.009  	&-0.2454 	&-0.2101 	\\
l & (log $\theta$)(log $g$)        	& 0.5137 	&-1.5053    	& 1.9031 	\\
m & (log $\theta$)(${\rm [Fe/H]}$) 	&-1.2216 	&-2.6334 	&-2.4474 	\\
n & (log $\theta$)([$\alpha$/Fe])  	& 0      	&-2.4218 	&-1.7837 	\\
o & (log $\theta$)$^2$(log $g$)     	& 2.7040 	& 0       	&-4.0214 	\\
p & (log $\theta$)(log $g$)$^2$     	& 0.2346 	& 0      	& 0      	\\
\hline
 & r.m.s. & 0.25 	& 0.21 	& 0.09	\\
\enddata
\end{deluxetable}

\begin{deluxetable}{lcc}
\tablenum{6} \tablewidth{0pt}
\tablecaption{Zero-points constants to be applied to indices derived from the fitting functions in order to calibrate them to the Lick/IDS system.  }
\label{calibration}
\tablehead{
Index   &   Zero-point & $\sigma$}
\startdata
Mg$_2$     &  0.018  & 0.035 \\
Fe5270  & -0.40  & 0.53 \\
Fe5335  & -0.66  & 0.53 \\
\enddata
\end{deluxetable}

%%%%%%%%%%%%%%%%%%%%%%%%%%%%%%%%%%%%%%%%%%%%%%%%%%%%%%%%%%%%%%%%%%%%
%%%%%%%%%%%%%%%%%%%%%%%%%%%%%%%%%%%%%%%%%%%%%%%%%%%%%%%%%%%%%%%%%%%%
%
%Table 7 Stellar population parameters
%
%%%%%%%%%%%%%%%%%%%%%%%%%%%%%%%%%%%%%%%%%%%%%%%%%%%%%%%%%%%%%%%%%%%%
%%%%%%%%%%%%%%%%%%%%%%%%%%%%%%%%%%%%%%%%%%%%%%%%%%%%%%%%%%%%%%%%%%%%

\begin{deluxetable}{lcccccc}
\tablenum{7} \tablewidth{0pt} \tablecaption{Parameters for the
best SSP fits to the compact group galaxies.} \tablecolumns{7}
\tablehead{
 & \multicolumn{3}{c}{TMB SSP models} & \multicolumn{3}{c}{This work SSP models} \\
 \colhead{ID} & \colhead{Age} & \colhead{$[Z/H]$} & \colhead{$[\alpha/Fe]$}
 & \colhead{$[Z/H]$} & \colhead{$[Fe/H]$} & \colhead{$[\alpha/Fe]$}
 }
 
\startdata
HCG 05B & 6 &  0.40 &  0.15 & 0.38 & 0.09 & 0.4 \\
HCG 21C &19 & -0.03 &  0.21 & -0.08& -0.34& 0.3 \\
HCG 76C &13 &  0.23 &  0.24 & 0.11 & -0.13& 0.3 \\
HCG 76D &13 &  0.07 &  0.45 & 0.02 & -0.35& 0.4 \\
HCG 86A &13 &  0.42 &  0.17 & 0.38 &  0.17& 0.3 \\
HCG 90B &11 &  0.36 &  0.30 & 0.25 & -0.06& 0.4 \\
HCG 90C &12 &  0.07 &  0.28 & 0.08 & -0.25& 0.4 \\
HCG 94A &17 &  0.39 &  0.25 & 0.20 & -0.01 & 0.2 \\
\enddata
\end{deluxetable}
\clearpage

\appendix

\section {Single stellar population model indices}

Single stellar population models described in Section 3.3.

\begin{deluxetable}{ccccccc}
\tablewidth{0pt}
\tablenum{A.1}
\tablecaption{SSP models indices for Mg$_2$, Fe5270 and Fe5335.} \label{sspmodels} 
\tablehead{
Age (Gyr)	&	Z	&	$[\alpha/Fe]$	&	$[Fe/H]$	&	Mg$_2$	&	Fe52	&	Fe53}
\startdata
4	&	0.008	&	0.0	&	-0.37	&	0.141	&	1.992	&	1.694	\\
5	&	0.008	&	0.0	&	-0.37	&	0.149	&	2.060	&	1.777	\\
6	&	0.008	&	0.0	&	-0.37	&	0.157	&	2.150	&	1.882	\\
7	&	0.008	&	0.0	&	-0.37	&	0.165	&	2.218	&	1.961	\\
8	&	0.008	&	0.0	&	-0.37	&	0.171	&	2.278	&	2.033	\\
9	&	0.008	&	0.0	&	-0.37	&	0.178	&	2.338	&	2.102	\\
10	&	0.008	&	0.0	&	-0.37	&	0.183	&	2.379	&	2.153	\\
11	&	0.008	&	0.0	&	-0.37	&	0.188	&	2.428	&	2.208	\\
12	&	0.008	&	0.0	&	-0.37	&	0.194	&	2.480	&	2.269	\\
13	&	0.008	&	0.0	&	-0.37	&	0.198	&	2.519	&	2.313	\\
14	&	0.008	&	0.0	&	-0.37	&	0.201	&	2.538	&	2.336	\\
15	&	0.008	&	0.0	&	-0.37	&	0.206	&	2.579	&	2.385	\\
16	&	0.008	&	0.0	&	-0.37	&	0.210	&	2.617	&	2.429	\\
4	&	0.008	&	0.2	&	-0.51	&	0.151	&	1.861	&	1.527	\\
5	&	0.008	&	0.2	&	-0.51	&	0.157	&	1.911	&	1.585	\\
6	&	0.008	&	0.2	&	-0.51	&	0.166	&	1.988	&	1.671	\\
7	&	0.008	&	0.2	&	-0.51	&	0.175	&	2.063	&	1.758	\\
8	&	0.008	&	0.2	&	-0.51	&	0.181	&	2.110	&	1.812	\\
9	&	0.008	&	0.2	&	-0.51	&	0.191	&	2.186	&	1.899	\\
10	&	0.008	&	0.2	&	-0.51	&	0.194	&	2.208	&	1.925	\\
11	&	0.008	&	0.2	&	-0.51	&	0.201	&	2.263	&	1.988	\\
12	&	0.008	&	0.2	&	-0.51	&	0.206	&	2.308	&	2.038	\\
13	&	0.008	&	0.2	&	-0.51	&	0.212	&	2.354	&	2.091	\\
14	&	0.008	&	0.2	&	-0.51	&	0.217	&	2.393	&	2.136	\\
15	&	0.008	&	0.2	&	-0.51	&	0.222	&	2.431	&	2.180	\\
16	&	0.008	&	0.2	&	-0.51	&	0.231	&	2.501	&	2.262	\\
4	&	0.008	&	0.4	&	-0.67	&	0.158	&	1.696	&	1.331	\\
5	&	0.008	&	0.4	&	-0.67	&	0.164	&	1.732	&	1.371	\\
6	&	0.008	&	0.4	&	-0.67	&	0.172	&	1.799	&	1.443	\\
7	&	0.008	&	0.4	&	-0.67	&	0.183	&	1.872	&	1.526	\\
8	&	0.008	&	0.4	&	-0.67	&	0.189	&	1.917	&	1.575	\\
9	&	0.008	&	0.4	&	-0.67	&	0.199	&	1.990	&	1.657	\\
10	&	0.008	&	0.4	&	-0.67	&	0.204	&	2.020	&	1.691	\\
11	&	0.008	&	0.4	&	-0.67	&	0.212	&	2.073	&	1.751	\\
12	&	0.008	&	0.4	&	-0.67	&	0.218	&	2.118	&	1.800	\\
13	&	0.008	&	0.4	&	-0.67	&	0.225	&	2.172	&	1.861	\\
14	&	0.008	&	0.4	&	-0.67	&	0.233	&	2.225	&	1.921	\\
15	&	0.008	&	0.4	&	-0.67	&	0.238	&	2.262	&	1.962	\\
4	&	0.01	&	0.0	&	-0.27	&	0.155	&	2.167	&	1.881	\\
5	&	0.01	&	0.0	&	-0.27	&	0.163	&	2.243	&	1.973	\\
6	&	0.01	&	0.0	&	-0.27	&	0.172	&	2.325	&	2.071	\\
7	&	0.01	&	0.0	&	-0.27	&	0.182	&	2.418	&	2.180	\\
8	&	0.01	&	0.0	&	-0.27	&	0.185	&	2.432	&	2.200	\\
9	&	0.01	&	0.0	&	-0.27	&	0.193	&	2.500	&	2.280	\\
10	&	0.01	&	0.0	&	-0.27	&	0.200	&	2.566	&	2.359	\\
11	&	0.01	&	0.0	&	-0.27	&	0.203	&	2.586	&	2.383	\\
12	&	0.01	&	0.0	&	-0.27	&	0.210	&	2.645	&	2.450	\\
13	&	0.01	&	0.0	&	-0.27	&	0.216	&	2.703	&	2.517	\\
14	&	0.01	&	0.0	&	-0.27	&	0.222	&	2.748	&	2.571	\\
15	&	0.01	&	0.0	&	-0.27	&	0.225	&	2.766	&	2.595	\\
16	&	0.01	&	0.0	&	-0.27	&	0.229	&	2.802	&	2.637	\\
4	&	0.01	&	0.2	&	-0.41	&	0.164	&	2.024	&	1.695	\\
5	&	0.01	&	0.2	&	-0.41	&	0.172	&	2.094	&	1.777	\\
6	&	0.01	&	0.2	&	-0.41	&	0.181	&	2.156	&	1.848	\\
7	&	0.01	&	0.2	&	-0.41	&	0.192	&	2.246	&	1.951	\\
8	&	0.01	&	0.2	&	-0.41	&	0.195	&	2.260	&	1.970	\\
9	&	0.01	&	0.2	&	-0.41	&	0.203	&	2.322	&	2.041	\\
10	&	0.01	&	0.2	&	-0.41	&	0.210	&	2.376	&	2.102	\\
11	&	0.01	&	0.2	&	-0.41	&	0.214	&	2.407	&	2.140	\\
12	&	0.01	&	0.2	&	-0.41	&	0.222	&	2.468	&	2.210	\\
13	&	0.01	&	0.2	&	-0.41	&	0.229	&	2.522	&	2.270	\\
14	&	0.01	&	0.2	&	-0.41	&	0.236	&	2.576	&	2.334	\\
15	&	0.01	&	0.2	&	-0.41	&	0.237	&	2.574	&	2.330	\\
16	&	0.01	&	0.2	&	-0.41	&	0.246	&	2.645	&	2.414	\\
4	&	0.01	&	0.4	&	-0.57	&	0.171	&	1.850	&	1.485	\\
5	&	0.01	&	0.4	&	-0.57	&	0.181	&	1.922	&	1.567	\\
6	&	0.01	&	0.4	&	-0.57	&	0.187	&	1.962	&	1.610	\\
7	&	0.01	&	0.4	&	-0.57	&	0.200	&	2.048	&	1.707	\\
8	&	0.01	&	0.4	&	-0.57	&	0.203	&	2.064	&	1.725	\\
9	&	0.01	&	0.4	&	-0.57	&	0.212	&	2.129	&	1.798	\\
10	&	0.01	&	0.4	&	-0.57	&	0.220	&	2.183	&	1.858	\\
11	&	0.01	&	0.4	&	-0.57	&	0.226	&	2.219	&	1.900	\\
12	&	0.01	&	0.4	&	-0.57	&	0.234	&	2.278	&	1.967	\\
13	&	0.01	&	0.4	&	-0.57	&	0.241	&	2.327	&	2.021	\\
14	&	0.01	&	0.4	&	-0.57	&	0.250	&	2.397	&	2.102	\\
15	&	0.01	&	0.4	&	-0.57	&	0.252	&	2.402	&	2.105	\\
16	&	0.01	&	0.4	&	-0.57	&	0.262	&	2.475	&	2.190	\\
4	&	0.014	&	0.0	&	-0.12	&	0.175	&	2.441	&	2.174	\\
5	&	0.014	&	0.0	&	-0.12	&	0.185	&	2.525	&	2.277	\\
6	&	0.014	&	0.0	&	-0.12	&	0.193	&	2.593	&	2.358	\\
7	&	0.014	&	0.0	&	-0.12	&	0.203	&	2.670	&	2.453	\\
8	&	0.014	&	0.0	&	-0.12	&	0.206	&	2.692	&	2.482	\\
9	&	0.014	&	0.0	&	-0.12	&	0.214	&	2.754	&	2.559	\\
10	&	0.014	&	0.0	&	-0.12	&	0.222	&	2.826	&	2.646	\\
11	&	0.014	&	0.0	&	-0.12	&	0.228	&	2.860	&	2.685	\\
12	&	0.014	&	0.0	&	-0.12	&	0.235	&	2.928	&	2.768	\\
13	&	0.014	&	0.0	&	-0.12	&	0.242	&	2.979	&	2.827	\\
14	&	0.014	&	0.0	&	-0.12	&	0.249	&	3.038	&	2.899	\\
15	&	0.014	&	0.0	&	-0.12	&	0.253	&	3.059	&	2.928	\\
16	&	0.014	&	0.0	&	-0.12	&	0.259	&	3.103	&	2.980	\\
4	&	0.014	&	0.2	&	-0.26	&	0.182	&	2.265	&	1.944	\\
5	&	0.014	&	0.2	&	-0.26	&	0.192	&	2.341	&	2.034	\\
6	&	0.014	&	0.2	&	-0.26	&	0.201	&	2.399	&	2.101	\\
7	&	0.014	&	0.2	&	-0.26	&	0.210	&	2.466	&	2.179	\\
8	&	0.014	&	0.2	&	-0.26	&	0.215	&	2.492	&	2.213	\\
9	&	0.014	&	0.2	&	-0.26	&	0.221	&	2.537	&	2.264	\\
10	&	0.014	&	0.2	&	-0.26	&	0.231	&	2.614	&	2.353	\\
11	&	0.014	&	0.2	&	-0.26	&	0.237	&	2.652	&	2.399	\\
12	&	0.014	&	0.2	&	-0.26	&	0.246	&	2.721	&	2.480	\\
13	&	0.014	&	0.2	&	-0.26	&	0.253	&	2.768	&	2.536	\\
14	&	0.014	&	0.2	&	-0.26	&	0.260	&	2.819	&	2.594	\\
15	&	0.014	&	0.2	&	-0.26	&	0.262	&	2.826	&	2.598	\\
4	&	0.014	&	0.4	&	-0.42	&	0.189	&	2.080	&	1.716	\\
5	&	0.014	&	0.4	&	-0.42	&	0.201	&	2.161	&	1.809	\\
6	&	0.014	&	0.4	&	-0.42	&	0.209	&	2.199	&	1.849	\\
7	&	0.014	&	0.4	&	-0.42	&	0.219	&	2.260	&	1.919	\\
8	&	0.014	&	0.4	&	-0.42	&	0.224	&	2.291	&	1.956	\\
9	&	0.014	&	0.4	&	-0.42	&	0.231	&	2.333	&	2.003	\\
10	&	0.014	&	0.4	&	-0.42	&	0.241	&	2.410	&	2.090	\\
11	&	0.014	&	0.4	&	-0.42	&	0.248	&	2.449	&	2.133	\\
12	&	0.014	&	0.4	&	-0.42	&	0.258	&	2.518	&	2.214	\\
13	&	0.014	&	0.4	&	-0.42	&	0.265	&	2.559	&	2.260	\\
14	&	0.014	&	0.4	&	-0.42	&	0.273	&	2.619	&	2.328	\\
15	&	0.014	&	0.4	&	-0.42	&	0.275	&	2.618	&	2.325	\\
16	&	0.014	&	0.4	&	-0.42	&	0.284	&	2.681	&	2.397	\\
4	&	0.018	&	0.0	&	0.00	&	0.188	&	2.632	&	2.379	\\
5	&	0.018	&	0.0	&	0.00	&	0.200	&	2.732	&	2.502	\\
6	&	0.018	&	0.0	&	0.00	&	0.209	&	2.797	&	2.579	\\
7	&	0.018	&	0.0	&	0.00	&	0.216	&	2.847	&	2.644	\\
8	&	0.018	&	0.0	&	0.00	&	0.224	&	2.911	&	2.723	\\
9	&	0.018	&	0.0	&	0.00	&	0.230	&	2.955	&	2.782	\\
10	&	0.018	&	0.0	&	0.00	&	0.239	&	3.026	&	2.867	\\
11	&	0.018	&	0.0	&	0.00	&	0.247	&	3.087	&	2.945	\\
12	&	0.018	&	0.0	&	0.00	&	0.256	&	3.151	&	3.023	\\
13	&	0.018	&	0.0	&	0.00	&	0.261	&	3.190	&	3.070	\\
14	&	0.018	&	0.0	&	0.00	&	0.269	&	3.251	&	3.146	\\
15	&	0.018	&	0.0	&	0.00	&	0.276	&	3.295	&	3.201	\\
16	&	0.018	&	0.0	&	0.00	&	0.282	&	3.347	&	3.269	\\
4	&	0.018	&	0.2	&	-0.14	&	0.195	&	2.442	&	2.126	\\
5	&	0.018	&	0.2	&	-0.14	&	0.206	&	2.522	&	2.222	\\
6	&	0.018	&	0.2	&	-0.14	&	0.216	&	2.583	&	2.292	\\
7	&	0.018	&	0.2	&	-0.14	&	0.223	&	2.623	&	2.342	\\
8	&	0.018	&	0.2	&	-0.14	&	0.231	&	2.683	&	2.413	\\
9	&	0.018	&	0.2	&	-0.14	&	0.238	&	2.718	&	2.457	\\
10	&	0.018	&	0.2	&	-0.14	&	0.248	&	2.800	&	2.552	\\
11	&	0.018	&	0.2	&	-0.14	&	0.258	&	2.861	&	2.626	\\
12	&	0.018	&	0.2	&	-0.14	&	0.266	&	2.928	&	2.705	\\
13	&	0.018	&	0.2	&	-0.14	&	0.273	&	2.967	&	2.752	\\
14	&	0.018	&	0.2	&	-0.14	&	0.279	&	3.010	&	2.805	\\
15	&	0.018	&	0.2	&	-0.14	&	0.285	&	3.043	&	2.841	\\
16	&	0.018	&	0.2	&	-0.14	&	0.293	&	3.096	&	2.905	\\
4	&	0.018	&	0.4	&	-0.30	&	0.203	&	2.249	&	1.885	\\
5	&	0.018	&	0.4	&	-0.30	&	0.215	&	2.328	&	1.977	\\
6	&	0.018	&	0.4	&	-0.30	&	0.224	&	2.376	&	2.029	\\
7	&	0.018	&	0.4	&	-0.30	&	0.231	&	2.412	&	2.071	\\
8	&	0.018	&	0.4	&	-0.30	&	0.241	&	2.470	&	2.138	\\
9	&	0.018	&	0.4	&	-0.30	&	0.247	&	2.502	&	2.174	\\
10	&	0.018	&	0.4	&	-0.30	&	0.260	&	2.587	&	2.273	\\
11	&	0.018	&	0.4	&	-0.30	&	0.268	&	2.640	&	2.333	\\
12	&	0.018	&	0.4	&	-0.30	&	0.279	&	2.712	&	2.416	\\
13	&	0.018	&	0.4	&	-0.30	&	0.285	&	2.749	&	2.460	\\
14	&	0.018	&	0.4	&	-0.30	&	0.293	&	2.793	&	2.508	\\
15	&	0.018	&	0.4	&	-0.30	&	0.296	&	2.810	&	2.527	\\
16	&	0.018	&	0.4	&	-0.30	&	0.305	&	2.862	&	2.587	\\
4	&	0.022	&	0.0	&	0.09	&	0.198	&	2.785	&	2.540	\\
5	&	0.022	&	0.0	&	0.09	&	0.213	&	2.912	&	2.699	\\
6	&	0.022	&	0.0	&	0.09	&	0.222	&	2.974	&	2.776	\\
7	&	0.022	&	0.0	&	0.09	&	0.229	&	3.017	&	2.833	\\
8	&	0.022	&	0.0	&	0.09	&	0.240	&	3.101	&	2.940	\\
9	&	0.022	&	0.0	&	0.09	&	0.246	&	3.143	&	2.993	\\
10	&	0.022	&	0.0	&	0.09	&	0.256	&	3.209	&	3.078	\\
11	&	0.022	&	0.0	&	0.09	&	0.265	&	3.282	&	3.169	\\
12	&	0.022	&	0.0	&	0.09	&	0.274	&	3.346	&	3.248	\\
13	&	0.022	&	0.0	&	0.09	&	0.280	&	3.386	&	3.301	\\
14	&	0.022	&	0.0	&	0.09	&	0.287	&	3.441	&	3.368	\\
15	&	0.022	&	0.0	&	0.09	&	0.295	&	3.501	&	3.448	\\
16	&	0.022	&	0.0	&	0.09	&	0.302	&	3.552	&	3.509	\\
4	&	0.022	&	0.2	&	-0.05	&	0.206	&	2.585	&	2.272	\\
5	&	0.022	&	0.2	&	-0.05	&	0.219	&	2.683	&	2.391	\\
6	&	0.022	&	0.2	&	-0.05	&	0.229	&	2.744	&	2.463	\\
7	&	0.022	&	0.2	&	-0.05	&	0.235	&	2.774	&	2.502	\\
8	&	0.022	&	0.2	&	-0.05	&	0.246	&	2.850	&	2.593	\\
9	&	0.022	&	0.2	&	-0.05	&	0.253	&	2.889	&	2.641	\\
10	&	0.022	&	0.2	&	-0.05	&	0.264	&	2.964	&	2.733	\\
11	&	0.022	&	0.2	&	-0.05	&	0.274	&	3.039	&	2.822	\\
12	&	0.022	&	0.2	&	-0.05	&	0.284	&	3.102	&	2.897	\\
13	&	0.022	&	0.2	&	-0.05	&	0.290	&	3.140	&	2.947	\\
14	&	0.022	&	0.2	&	-0.05	&	0.297	&	3.180	&	2.991	\\
15	&	0.022	&	0.2	&	-0.05	&	0.305	&	3.232	&	3.056	\\
4	&	0.022	&	0.4	&	-0.21	&	0.213	&	2.379	&	2.012	\\
5	&	0.022	&	0.4	&	-0.21	&	0.227	&	2.471	&	2.121	\\
6	&	0.022	&	0.4	&	-0.21	&	0.237	&	2.526	&	2.183	\\
7	&	0.022	&	0.4	&	-0.21	&	0.244	&	2.551	&	2.213	\\
8	&	0.022	&	0.4	&	-0.21	&	0.255	&	2.618	&	2.290	\\
9	&	0.022	&	0.4	&	-0.21	&	0.261	&	2.653	&	2.332	\\
10	&	0.022	&	0.4	&	-0.21	&	0.274	&	2.733	&	2.424	\\
11	&	0.022	&	0.4	&	-0.21	&	0.284	&	2.798	&	2.499	\\
12	&	0.022	&	0.4	&	-0.21	&	0.295	&	2.862	&	2.574	\\
13	&	0.022	&	0.4	&	-0.21	&	0.303	&	2.901	&	2.622	\\
14	&	0.022	&	0.4	&	-0.21	&	0.309	&	2.937	&	2.662	\\
15	&	0.022	&	0.4	&	-0.21	&	0.316	&	2.978	&	2.707	\\
16	&	0.022	&	0.4	&	-0.21	&	0.323	&	3.017	&	2.755	\\
4	&	0.026	&	0.0	&	0.17	&	0.207	&	2.929	&	2.694	\\
5	&	0.026	&	0.0	&	0.17	&	0.226	&	3.080	&	2.885	\\
6	&	0.026	&	0.0	&	0.17	&	0.235	&	3.138	&	2.962	\\
7	&	0.026	&	0.0	&	0.17	&	0.244	&	3.195	&	3.036	\\
8	&	0.026	&	0.0	&	0.17	&	0.253	&	3.268	&	3.133	\\
9	&	0.026	&	0.0	&	0.17	&	0.263	&	3.328	&	3.202	\\
10	&	0.026	&	0.0	&	0.17	&	0.271	&	3.387	&	3.282	\\
11	&	0.026	&	0.0	&	0.17	&	0.282	&	3.462	&	3.375	\\
12	&	0.026	&	0.0	&	0.17	&	0.290	&	3.530	&	3.466	\\
13	&	0.026	&	0.0	&	0.17	&	0.298	&	3.577	&	3.525	\\
14	&	0.026	&	0.0	&	0.17	&	0.305	&	3.626	&	3.588	\\
15	&	0.026	&	0.0	&	0.17	&	0.313	&	3.683	&	3.658	\\
16	&	0.026	&	0.0	&	0.17	&	0.320	&	3.736	&	3.725	\\
4	&	0.026	&	0.2	&	0.03	&	0.214	&	2.711	&	2.400	\\
5	&	0.026	&	0.2	&	0.03	&	0.231	&	2.832	&	2.549	\\
6	&	0.026	&	0.2	&	0.03	&	0.241	&	2.887	&	2.619	\\
7	&	0.026	&	0.2	&	0.03	&	0.248	&	2.924	&	2.668	\\
8	&	0.026	&	0.2	&	0.03	&	0.259	&	2.997	&	2.752	\\
9	&	0.026	&	0.2	&	0.03	&	0.269	&	3.050	&	2.820	\\
10	&	0.026	&	0.2	&	0.03	&	0.278	&	3.114	&	2.897	\\
11	&	0.026	&	0.2	&	0.03	&	0.290	&	3.191	&	2.995	\\
12	&	0.026	&	0.2	&	0.03	&	0.299	&	3.251	&	3.063	\\
13	&	0.026	&	0.2	&	0.03	&	0.307	&	3.296	&	3.118	\\
14	&	0.026	&	0.2	&	0.03	&	0.313	&	3.332	&	3.167	\\
15	&	0.026	&	0.2	&	0.03	&	0.322	&	3.391	&	3.239	\\
16	&	0.026	&	0.2	&	0.03	&	0.328	&	3.432	&	3.288	\\
4	&	0.026	&	0.4	&	-0.13	&	0.220	&	2.488	&	2.120	\\
5	&	0.026	&	0.4	&	-0.13	&	0.239	&	2.603	&	2.256	\\
6	&	0.026	&	0.4	&	-0.13	&	0.249	&	2.660	&	2.322	\\
7	&	0.026	&	0.4	&	-0.13	&	0.256	&	2.685	&	2.351	\\
8	&	0.026	&	0.4	&	-0.13	&	0.266	&	2.746	&	2.422	\\
9	&	0.026	&	0.4	&	-0.13	&	0.276	&	2.797	&	2.484	\\
10	&	0.026	&	0.4	&	-0.13	&	0.286	&	2.858	&	2.557	\\
11	&	0.026	&	0.4	&	-0.13	&	0.298	&	2.927	&	2.635	\\
12	&	0.026	&	0.4	&	-0.13	&	0.307	&	2.983	&	2.703	\\
13	&	0.026	&	0.4	&	-0.13	&	0.316	&	3.026	&	2.754	\\
14	&	0.026	&	0.4	&	-0.13	&	0.324	&	3.063	&	2.799	\\
15	&	0.026	&	0.4	&	-0.13	&	0.332	&	3.118	&	2.864	\\
16	&	0.026	&	0.4	&	-0.13	&	0.339	&	3.151	&	2.899	\\
4	&	0.03	&	0.0	&	0.24	&	0.216	&	3.064	&	2.842	\\
5	&	0.03	&	0.0	&	0.24	&	0.238	&	3.232	&	3.054	\\
6	&	0.03	&	0.0	&	0.24	&	0.246	&	3.283	&	3.125	\\
7	&	0.03	&	0.0	&	0.24	&	0.257	&	3.354	&	3.217	\\
8	&	0.03	&	0.0	&	0.24	&	0.266	&	3.419	&	3.304	\\
9	&	0.03	&	0.0	&	0.24	&	0.278	&	3.500	&	3.405	\\
10	&	0.03	&	0.0	&	0.24	&	0.286	&	3.549	&	3.468	\\
11	&	0.03	&	0.0	&	0.24	&	0.297	&	3.627	&	3.567	\\
12	&	0.03	&	0.0	&	0.24	&	0.306	&	3.694	&	3.653	\\
13	&	0.03	&	0.0	&	0.24	&	0.315	&	3.749	&	3.722	\\
14	&	0.03	&	0.0	&	0.24	&	0.321	&	3.792	&	3.783	\\
15	&	0.03	&	0.0	&	0.24	&	0.330	&	3.849	&	3.854	\\
16	&	0.03	&	0.0	&	0.24	&	0.337	&	3.906	&	3.934	\\
4	&	0.03	&	0.2	&	0.10	&	0.221	&	2.823	&	2.516	\\
5	&	0.03	&	0.2	&	0.10	&	0.242	&	2.963	&	2.687	\\
6	&	0.03	&	0.2	&	0.10	&	0.252	&	3.013	&	2.754	\\
7	&	0.03	&	0.2	&	0.10	&	0.260	&	3.057	&	2.810	\\
8	&	0.03	&	0.2	&	0.10	&	0.271	&	3.128	&	2.898	\\
9	&	0.03	&	0.2	&	0.10	&	0.281	&	3.193	&	2.975	\\
10	&	0.03	&	0.2	&	0.10	&	0.291	&	3.248	&	3.043	\\
11	&	0.03	&	0.2	&	0.10	&	0.302	&	3.325	&	3.141	\\
12	&	0.03	&	0.2	&	0.10	&	0.313	&	3.384	&	3.212	\\
13	&	0.03	&	0.2	&	0.10	&	0.320	&	3.435	&	3.280	\\
14	&	0.03	&	0.2	&	0.10	&	0.328	&	3.467	&	3.321	\\
15	&	0.03	&	0.2	&	0.10	&	0.336	&	3.531	&	3.400	\\
4	&	0.03	&	0.4	&	-0.06	&	0.228	&	2.590	&	2.220	\\
5	&	0.03	&	0.4	&	-0.06	&	0.249	&	2.720	&	2.374	\\
6	&	0.03	&	0.4	&	-0.06	&	0.260	&	2.774	&	2.440	\\
7	&	0.03	&	0.4	&	-0.06	&	0.267	&	2.800	&	2.474	\\
8	&	0.03	&	0.4	&	-0.06	&	0.278	&	2.863	&	2.543	\\
9	&	0.03	&	0.4	&	-0.06	&	0.288	&	2.923	&	2.617	\\
10	&	0.03	&	0.4	&	-0.06	&	0.298	&	2.971	&	2.676	\\
11	&	0.03	&	0.4	&	-0.06	&	0.311	&	3.039	&	2.758	\\
12	&	0.03	&	0.4	&	-0.06	&	0.319	&	3.092	&	2.819	\\
13	&	0.03	&	0.4	&	-0.06	&	0.329	&	3.140	&	2.880	\\
14	&	0.03	&	0.4	&	-0.06	&	0.336	&	3.177	&	2.922	\\
15	&	0.03	&	0.4	&	-0.06	&	0.347	&	3.241	&	2.998	\\
16	&	0.03	&	0.4	&	-0.06	&	0.352	&	3.274	&	3.037	\\
4	&	0.035	&	0.0	&	0.32	&	0.226	&	3.223	&	3.012	\\
5	&	0.035	&	0.0	&	0.32	&	0.250	&	3.398	&	3.236	\\
6	&	0.035	&	0.0	&	0.32	&	0.259	&	3.447	&	3.311	\\
7	&	0.035	&	0.0	&	0.32	&	0.272	&	3.537	&	3.431	\\
8	&	0.035	&	0.0	&	0.32	&	0.280	&	3.592	&	3.506	\\
9	&	0.035	&	0.0	&	0.32	&	0.294	&	3.686	&	3.615	\\
10	&	0.035	&	0.0	&	0.32	&	0.303	&	3.736	&	3.683	\\
11	&	0.035	&	0.0	&	0.32	&	0.313	&	3.810	&	3.778	\\
12	&	0.035	&	0.0	&	0.32	&	0.324	&	3.880	&	3.869	\\
13	&	0.035	&	0.0	&	0.32	&	0.333	&	3.945	&	3.956	\\
14	&	0.035	&	0.0	&	0.32	&	0.340	&	3.979	&	4.003	\\
15	&	0.035	&	0.0	&	0.32	&	0.348	&	4.040	&	4.082	\\
16	&	0.035	&	0.0	&	0.32	&	0.357	&	4.094	&	4.151	\\
4	&	0.035	&	0.2	&	0.18	&	0.230	&	2.956	&	2.655	\\
5	&	0.035	&	0.2	&	0.18	&	0.254	&	3.107	&	2.841	\\
6	&	0.035	&	0.2	&	0.18	&	0.263	&	3.155	&	2.906	\\
7	&	0.035	&	0.2	&	0.18	&	0.272	&	3.206	&	2.972	\\
8	&	0.035	&	0.2	&	0.18	&	0.283	&	3.275	&	3.058	\\
9	&	0.035	&	0.2	&	0.18	&	0.296	&	3.350	&	3.148	\\
10	&	0.035	&	0.2	&	0.18	&	0.305	&	3.399	&	3.214	\\
11	&	0.035	&	0.2	&	0.18	&	0.317	&	3.475	&	3.307	\\
12	&	0.035	&	0.2	&	0.18	&	0.327	&	3.535	&	3.385	\\
13	&	0.035	&	0.2	&	0.18	&	0.337	&	3.591	&	3.455	\\
14	&	0.035	&	0.2	&	0.18	&	0.342	&	3.621	&	3.495	\\
15	&	0.035	&	0.2	&	0.18	&	0.353	&	3.688	&	3.585	\\
16	&	0.035	&	0.2	&	0.18	&	0.359	&	3.730	&	3.630	\\
4	&	0.035	&	0.4	&	0.02	&	0.237	&	2.705	&	2.333	\\
5	&	0.035	&	0.4	&	0.02	&	0.259	&	2.845	&	2.502	\\
6	&	0.035	&	0.4	&	0.02	&	0.270	&	2.898	&	2.569	\\
7	&	0.035	&	0.4	&	0.02	&	0.278	&	2.928	&	2.606	\\
8	&	0.035	&	0.4	&	0.02	&	0.291	&	2.993	&	2.680	\\
9	&	0.035	&	0.4	&	0.02	&	0.303	&	3.060	&	2.762	\\
10	&	0.035	&	0.4	&	0.02	&	0.312	&	3.097	&	2.811	\\
11	&	0.035	&	0.4	&	0.02	&	0.323	&	3.164	&	2.891	\\
12	&	0.035	&	0.4	&	0.02	&	0.334	&	3.216	&	2.950	\\
13	&	0.035	&	0.4	&	0.02	&	0.342	&	3.271	&	3.019	\\
14	&	0.035	&	0.4	&	0.02	&	0.352	&	3.310	&	3.063	\\
15	&	0.035	&	0.4	&	0.02	&	0.361	&	3.376	&	3.149	\\
4	&	0.04	&	0.0	&	0.39	&	0.236	&	3.372	&	3.175	\\
5	&	0.04	&	0.0	&	0.39	&	0.261	&	3.542	&	3.398	\\
6	&	0.04	&	0.0	&	0.39	&	0.271	&	3.594	&	3.477	\\
7	&	0.04	&	0.0	&	0.39	&	0.286	&	3.694	&	3.608	\\
8	&	0.04	&	0.0	&	0.39	&	0.294	&	3.746	&	3.681	\\
9	&	0.04	&	0.0	&	0.39	&	0.309	&	3.847	&	3.796	\\
10	&	0.04	&	0.0	&	0.39	&	0.317	&	3.898	&	3.870	\\
11	&	0.04	&	0.0	&	0.39	&	0.329	&	3.973	&	3.966	\\
12	&	0.04	&	0.0	&	0.39	&	0.339	&	4.048	&	4.068	\\
13	&	0.04	&	0.0	&	0.39	&	0.350	&	4.113	&	4.152	\\
14	&	0.04	&	0.0	&	0.39	&	0.355	&	4.149	&	4.203	\\
15	&	0.04	&	0.0	&	0.39	&	0.365	&	4.207	&	4.278	\\
16	&	0.04	&	0.0	&	0.39	&	0.372	&	4.259	&	4.344	\\
4	&	0.04	&	0.2	&	0.25	&	0.239	&	3.080	&	2.786	\\
5	&	0.04	&	0.2	&	0.25	&	0.263	&	3.232	&	2.974	\\
6	&	0.04	&	0.2	&	0.25	&	0.274	&	3.284	&	3.048	\\
7	&	0.04	&	0.2	&	0.25	&	0.284	&	3.340	&	3.121	\\
8	&	0.04	&	0.2	&	0.25	&	0.295	&	3.404	&	3.199	\\
9	&	0.04	&	0.2	&	0.25	&	0.310	&	3.486	&	3.301	\\
10	&	0.04	&	0.2	&	0.25	&	0.318	&	3.533	&	3.362	\\
11	&	0.04	&	0.2	&	0.25	&	0.331	&	3.610	&	3.463	\\
12	&	0.04	&	0.2	&	0.25	&	0.341	&	3.669	&	3.535	\\
13	&	0.04	&	0.2	&	0.25	&	0.352	&	3.729	&	3.611	\\
14	&	0.04	&	0.2	&	0.25	&	0.358	&	3.762	&	3.661	\\
15	&	0.04	&	0.2	&	0.25	&	0.369	&	3.828	&	3.738	\\
16	&	0.04	&	0.2	&	0.25	&	0.375	&	3.875	&	3.803	\\
4	&	0.04	&	0.4	&	0.09	&	0.245	&	2.813	&	2.442	\\
5	&	0.04	&	0.4	&	0.09	&	0.268	&	2.955	&	2.614	\\
6	&	0.04	&	0.4	&	0.09	&	0.280	&	3.010	&	2.685	\\
7	&	0.04	&	0.4	&	0.09	&	0.289	&	3.044	&	2.727	\\
8	&	0.04	&	0.4	&	0.09	&	0.301	&	3.108	&	2.801	\\
9	&	0.04	&	0.4	&	0.09	&	0.315	&	3.179	&	2.891	\\
10	&	0.04	&	0.4	&	0.09	&	0.323	&	3.213	&	2.934	\\
11	&	0.04	&	0.4	&	0.09	&	0.336	&	3.280	&	3.014	\\
12	&	0.04	&	0.4	&	0.09	&	0.344	&	3.330	&	3.077	\\
13	&	0.04	&	0.4	&	0.09	&	0.356	&	3.390	&	3.146	\\
14	&	0.04	&	0.4	&	0.09	&	0.363	&	3.431	&	3.200	\\
15	&	0.04	&	0.4	&	0.09	&	0.375	&	3.497	&	3.282	\\
\enddata
\end{deluxetable}

\end{document}